\documentclass{aa}  

\usepackage{graphicx}
\usepackage{txfonts}
\usepackage{url}
\usepackage{hyperref}
\hypersetup{
     breaklinks,
     colorlinks   = true,
     citecolor    = blue,
     urlcolor     = blue,
     linkcolor    = blue
}

\usepackage{placeins}
\usepackage[rightcaption]{sidecap}

\begin{document} 

   \title{Limitations and rotation of the two-armed phase spiral\\ in the Milky Way stellar disc}

   \author{S. Alinder\inst{1},
          P. J. McMillan\inst{2}
          \and
          T. Bensby \inst{1}
          }

   \institute{Lund Observatory, Division of Astrophysics, Department of Physics, 
            Lund University, Box 118, SE-221\,00 Lund, Sweden \and
         School of Physics \& Astronomy, University of Leicester, University Road, Leicester LE1 7RH, UK.\\
              \email{simon.alinder@fysik.lu.se}
             }

   \date{Received 5 July 2024; accepted YYYY}
 
  \abstract
   {
    The Milky Way's history of recent disturbances is vividly demonstrated by a structure in the vertical phase-space distribution known as the Gaia phase spiral.
    A one-armed phase spiral has been seen widely across the  Milky Way disc, while a two-armed one has only been observed in the solar neighbourhood.
   }
   {
   This study aims to determine the properties of the two-armed phase spiral and to put it in a Galactic context, with the ultimate goal of understanding the structure and history of the Milky Way disc.
   }
   {
    The {\sl Gaia} DR3 data is used to trace and characterise the two-armed phase spiral. Special focus is put on the phase spiral's spatial distribution, rotational behaviour, and chemical characteristics. To quantify the properties of the phase spiral
    we use a model that fits a spiral pattern to the phase space distribution of the stars.
   }
   {
    We find that the two-armed phase spiral is detectable only within a narrow range of galactocentric distances and angular momenta in the solar neighbourhood, $R = 8 \pm 0.5 $ kpc, $L_Z = 1450 \pm 50$ kpc km s$^{-1}$.
    Outside this region, the phase spiral is one-armed.
    The two-armed phase spiral rotates with the phase angle, like the one-armed phase spiral, and changes axis ratio with phase angle.
    Additionally, stars within the phase-space overdensity caused by the two-armed phase spiral pattern have slightly higher mean metallicity than stars in the underdense regions of the pattern at equivalent galactocentric distances, angular momenta, and vertical orbit extents.
   }
  % conclusions heading (optional)
   {
     The two-armed phase spiral rotates with phase angle and its effect can be seen in metallicity, like the one-armed phase spiral.
     However, the limited range over which it can be found, and its variation in shape are quite different from the one-armed version, suggesting it is a much more localised phenomenon in the Galactic disc.
     %The detectability of the two-armed phase spiral in mean metallicity data suggests chemical distinctions among stars participating in the phase spiral compared to stars on similar orbits that are not part of the phase spiral.
   }


   \maketitle

%-------------------------------------------------------------------

\section{Introduction}
The processes that reshape galaxies like the Milky Way, such as mergers \citep[e.g.,][]{white_core_1978} and the creation and influence of bars and spiral arms \citep[e.g.,][]{lindblad_development_1941, sellwood_radial_2002} cause disequilibrium within galaxies. Studying this disequilibrium in the Milky Way gives us the best chance of understanding how these processes operate.

The quantity and quality of astrometric and spectroscopic data from the European Space Agency's \textit{Gaia} mission \citep{gaia_collaboration_gaia_2016, gaia_collaboration_gaia_2016-1, gaia_collaboration_gaia_2018, gaia_collaboration_gaia_2021, gaia_collaboration_gaia_2023}, has revolutionised our access to the Milky Way's structure in recent years and has allowed astronomers to study our Galaxy in greater detail than ever before.
\cite{antoja_dynamically_2018} discovered the ``phase spiral'' or ``snail shell'' in the velocity-space distribution of stars in the second \textit{Gaia} data release. 
The phase spiral is an unevenness in the distribution of vertical velocities of stars in the Galactic disc, which takes the form of a spiral when viewed in a vertical phase space diagram with $Z$ and $V_Z$ on the axes.
Subsequent investigations have expanded upon these findings, exploring the phase spiral's extent, morphology, and dependence on age or metallicity.
These have shown that the phase spiral covers a smaller range in vertical velocity with increasing galactocentric radius \citep[e.g.,][]{laporte_footprints_2019, antoja_phase_2023}, rotates with position around the galaxy  \citep[e.g.,][]{alinder_investigating_2023}, and has higher contrast in the inner regions for younger, more metal-rich stars \citep{bland-hawthorn_galah_2019}.

The origin of the phase spiral is a subject of active discussion. Suggested theories include the impact of the Sagittarius dwarf galaxy's passage through the Milky Way disc \citep{binney_origin_2018, laporte_footprints_2019, bland-hawthorn_galah_2019}, % These are the exact same citations as in paper 1.
and waves induced by a change in the pattern speed of the bar \citep{li_gaia_2023}.
The possibility that there are multiple causes behind the observed phenomenon has emerged in recent years. 
\cite{hunt_multiple_2022, bennett_exploring_2022} and \cite{tremaine_origin_2023} all suggest that the formation history of the phase spiral cannot be explained with a single impact but perhaps from several smaller disturbances.
\cite{frankel_vertical_2023} and \cite{antoja_phase_2023} both find that a simple model with a single cause for the perturbation fails to explain the observations and calls for more complex models.

The first mention of two-armed phase spirals in the literature is in the N-body merger simulations of \cite{hunt_resolving_2021} where they show up as the result of an interaction with an external perturber.
These two-armed spirals are relatively short-lived, but the authors claim that traces of them might still be found in the outer Galactic disc which preserves dynamical structures for longer.
In \cite{hunt_multiple_2022}, however, the two-armed phase spiral was seen in \textit{Gaia} data release 3 (DR3) in solar neighbourhood stars (with heliocentric distances less than 1 kpc) with a guiding centre radius ($R_G$) of about $6.33\,\mathrm{kpc}$.
The concept of the guiding centre comes from epicycle theory, in which the orbital motion of a star can be decomposed into a guiding centre on a large circular orbit with the same angular momentum as the star, and a small epicyclic elliptical excursion around this centre.
Even for substantially non-circular orbits, we can still use this concept as a convenient description via action-angle coordinates \citep[e.g.,][]{binney_galactic_2008}, with the position of the guiding centre being associated with the azimuthal action (that is, the angular momentum) and its conjugate angle. 
\cite{hunt_multiple_2022} investigated the phase spiral as a function of $R_G$ and azimuthal angle of the guiding centre ($\theta_\phi$) and detected phase spirals with both one and two arms.

Using an $N$-body simulation of an isolated Milky Way-like galaxy, \cite{hunt_multiple_2022} found that this sort of system forms a two-armed phase spiral in the inner regions of the galaxy, but no one-armed phase spiral. 
\cite{grand_ever-present_2023} conduct a cosmological simulation of a Milky Way-like galaxy experiencing an interaction with an external, Sagittarius-like, perturber. 
This simulation shows transient two-armed phase spirals appearing in different regions of the galaxy. 
\cite{li_gaia_2023} investigate how an isolated spiral galaxy can form phase spirals. 
Central to their analysis is the concept of ``pattern speed'', defined here as the rate at which the structure of the bar rotates. 
They find that an isolated spiral galaxy containing a bar with constant pattern speed does not form any phase spirals, despite producing a vertical perturbation in the galaxy's disc, called a ``breathing mode'', which is usually associated with two-armed phase spirals \citep{widrow_swing_2023}.
When they introduce a bar with a decreasing pattern speed, a two-armed phase spiral is created. 
They claim that a galaxy with internal non-axisymmetric structures with a constant pattern speed will not produce the observed structures in vertical motion when encountering an external perturber. % chemical composition?

This paper aims to address some key questions about the characteristics of the two-armed phase spiral in the Milky Way.
Specifically, we seek to determine where in the Galaxy and where in phase space this two-armed phase spiral can be found, how it varies with different parameters such as azimuthal angle ($\phi$), phase angle ($\theta_\phi$), and angular momentum ($L_Z$). We will further investigate whether it has any characteristic appearance in metallicity space.
It is our belief that, by establishing the behaviour of the two-armed phase spiral and comparing it to that of the one-armed phase spiral, we can gain insights into the dynamical processes shaping our Galaxy and the interplay between these processes and the Galactic chemical environment. 

We begin by describing the selection and analysis of the stellar sample in Sect.~\ref{section:data}.
In Sect.~\ref{section:results} we present the methods we utilise in searching for and characterising the two-armed phase spiral, and the results.
Our interpretation and discussion of the results are presented in Sect.~\ref{section:discussion}, where we put our findings into a broader context.
Finally, Sect.~\ref{section:conclusions} summarises our key conclusions and the implications they may have for our understanding of the assembly and evolution of our Galaxy.

\section{Data}\label{section:data}
\begin{figure} 
\centering
\includegraphics[width=\hsize]{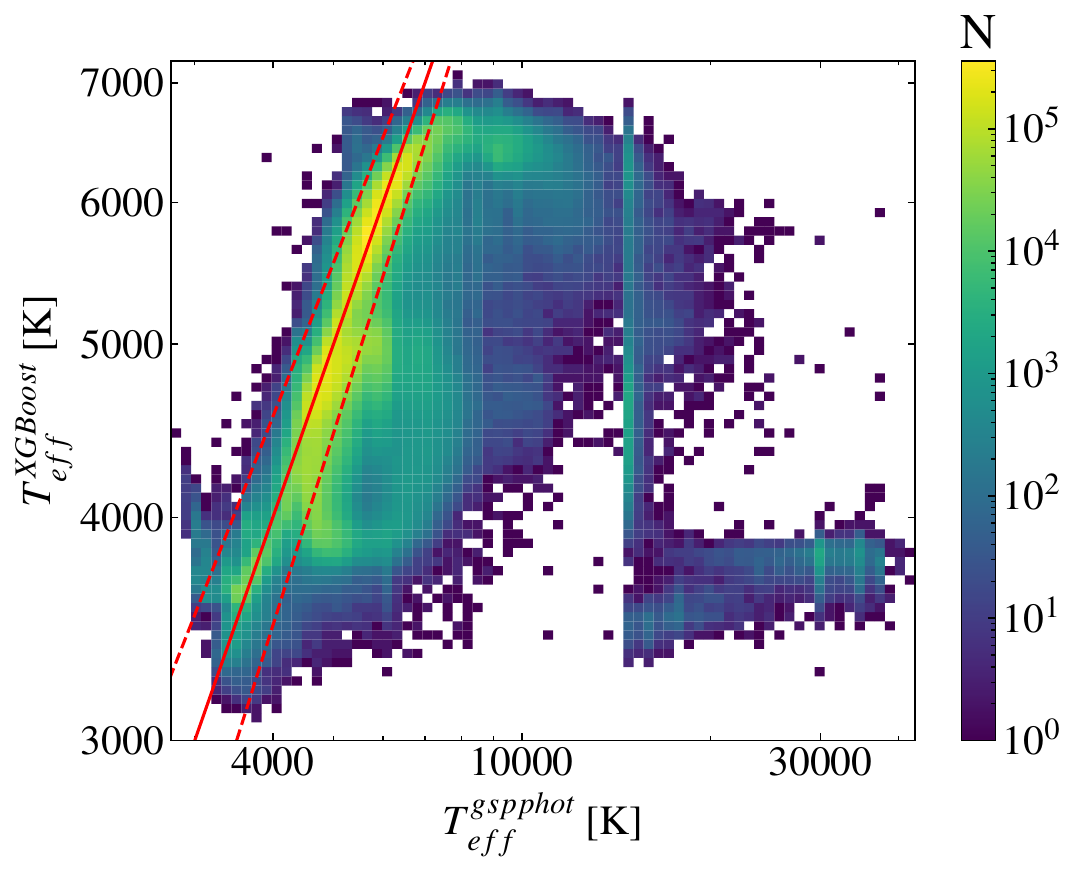}
  \caption{Comparison of $T_{\mathrm{eff}}^{gspphot} $ and $ T_{\mathrm{eff}}^{xgboost}$. The solid red line is the line of equality. The dashed red lines mark $|T_{\mathrm{eff}}^{gspphot} - T_{\mathrm{eff}}^{xgboost}| < 500 \,\mathrm{K}$ which is the selection we use. 
          }
     \label{fig:teff_diff}
\end{figure}

\subsection{Astrometic data}
We use a standard coordinate system\footnote{
	The galactocentric coordinate system used in this study places the Sun on the negative $X$-axis ($\phi = 180^\circ$) at a distance of 8.122 kpc and a height of 20.8 pc, with the $Y$-axis in the same direction as $l=90^{\circ}$, and the $Z$-axis is in the same direction as $b=90^{\circ}$. Galactic azimuth, $\phi$, decreases in the direction of Galactic rotation, and the Sun has velocity components $V_{R, \sun}=-12.9, V_{\phi, \sun}=-245.6, V_{Z, \sun}=-7.78$ km\,s$^{-1}$.
	\citep{reid_proper_2004, drimmel_solar_2018, gravity_collaboration_detection_2018, bennett_vertical_2019}. For the computations and definitions of coordinates, we use  Astropy v5.2 \citep{astropy_collaboration_astropy_2022}.
} and \textit{Gaia} DR3 to get the position and velocities \citep{gaia_collaboration_gaia_2016, gaia_collaboration_gaia_2023} of stars, with distances calculated by \cite{bailer-jones_estimating_2021}, who used a Bayesian approach with a direction-dependant prior on distance, the measured parallaxes, \textit{Gaia} photometry, and the fact that stars of different colours have different ranges of probable absolute magnitudes to re-compute the distances.

% Astropy references for latest Galactocentric parameters. 
%'galcen_coord': 'https://ui.adsabs.harvard.edu/abs/2004ApJ...616..872R',
% 'galcen_distance': 'https://ui.adsabs.harvard.edu/abs/2018A%26A...615L..15G',
% 'galcen_v_sun': ['https://ui.adsabs.harvard.edu/abs/2018RNAAS...2..210D',
%  'https://ui.adsabs.harvard.edu/abs/2018A%26A...615L..15G',
%  'https://ui.adsabs.harvard.edu/abs/2004ApJ...616..872R'],
% 'z_sun': 'https://ui.adsabs.harvard.edu/abs/2019MNRAS.482.1417B',
% 'roll': None}

When querying the public \textit{Gaia} database\footnote{\url{https://gea.esac.esa.int/archive/}}, we require \verb|parallax_over_error>=3 | as this removes the most uncertain distance estimates.
The query used resulted in 31\,552\,449 stars being selected.
The full ADQL-query used to retrieve this data was:
\begin{verbatim}
SELECT source_id, ra, dec, pmra, pmdec, 
r_med_photogeo, radial_velocity,
bp_rp, pmra_error, pmdec_error,
parallax_error, teff_gspphot
FROM external.gaiaedr3_distance
JOIN gaiadr3.gaia_source USING (source_id)
WHERE parallax_over_error>=3 
and radial_velocity IS NOT NULL 
and r_med_photogeo IS NOT NULL
\end{verbatim}

\subsection{Metallicity data}

\sidecaptionvpos{figure}{c}
\begin{SCfigure*}[0.4]
\includegraphics[width=0.6\textwidth]{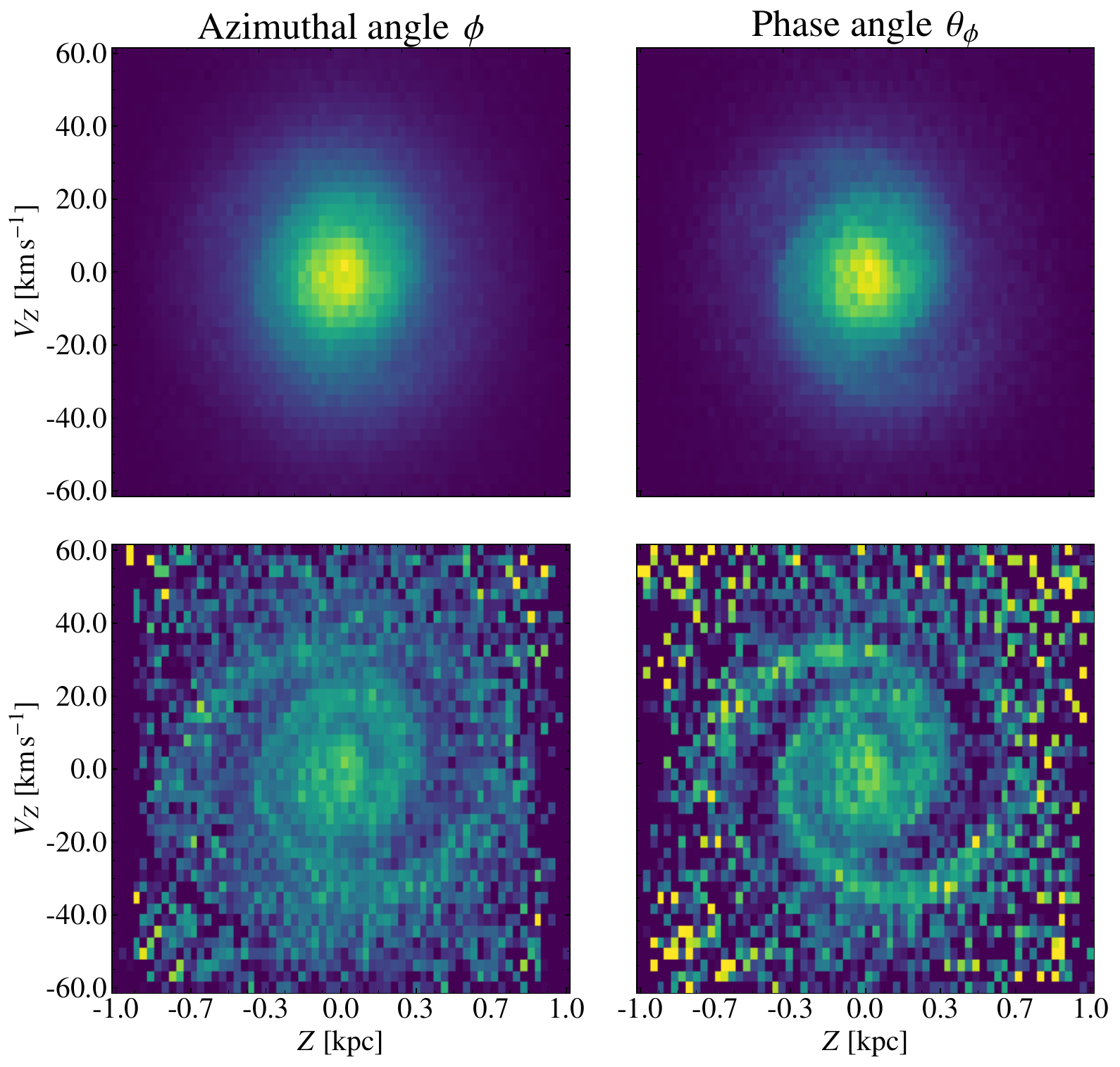}
  \caption{Comparison of phase spiral clarity when selecting stars based on azimuthal angle ($\phi$, \textit{left column}) and azimuthal angle of the guiding centre or phase angle ($\theta_\phi$, \textit{right column}), and using a Gaussian filter (\textit{bottom row}) compared to not using one (\textit{top row}).
  We can note that the right column, particularly the bottom row, has a much clearer spiral pattern than the rest.
  The plots show number density and number density contrast in the $Z-V_Z$ phase plane of stars in the solar neighbourhood ($d < 1$ kpc) with a further selection by $\phi$ or $\theta_\phi$. The left column shows stars with $175^\circ < \phi < 185^\circ$ and the right column shows stars with $175^\circ < \theta_\phi < 185^\circ$.
  All stars have $1400 < L_Z \,/ \,\mathrm{kpc\, km\, s}^{-1} < 1500$.
          }
     \label{fig:nearby_theta}
\end{SCfigure*}

\cite{andrae_robust_2023} used a machine learning algorithm called \verb|XGBoost| to derive stellar metallicities, $T_{\mathrm{eff}}$, and $\log(g)$ for more than 120 million stars in the \textit{Gaia} catalogue based on data from its low-resolution spectra. 
From this data, we select the 25\,050\,958 stars that are also present in our selection from the \textit{Gaia} database.

To address inaccuracies in classifying stars with high effective temperatures ($T_{\mathrm{eff}} > 7000 \, \mathrm{K}$), due to the lack of such stars in the \cite{andrae_robust_2023} training data, we apply additional cuts to the data.
To compensate for this, we compared the effective temperature within \textit{Gaia} DR3 (GSP-Phot) to that reported by XGBoost.
The discrepancy between the effective temperatures reported by \textit{Gaia}'s GSP-Phot and those predicted by the \verb|XGBoost| algorithm is highlighted in Fig. \ref{fig:teff_diff}, particularly for stars with $T_{\mathrm{eff}}^{gspphot} > 10\,000 \, \mathrm{K}$, where significant mismatches are observed.

To leave a clean sample we remove all stars that are listed as variables in the \textit{Gaia} catalogue because their temperature estimates are uncertain as $T_{\mathrm{eff}}$ varies with pulsation phase.
Then, all stars with a difference in effective temperature $|T_{\mathrm{eff}}^{gspphot} - T_{\mathrm{eff}}^{xgboost}| > 500 \,\mathrm{K}$ were removed.
Doing this cut also effectively removes all stars with $T_{\mathrm{eff}}^{gspphot} \gtrsim 7000 \,\mathrm{K}$ from our sample.
This leaves a sample of 15\,575\,635 stars with metallicity data.
The dashed red lines in Fig. \ref{fig:teff_diff} show this region.

\section{Characterising the two-armed phase spiral} \label{section:results}

\subsection{Data Processing}
% Explain the method used
% The two-armed spiral is much easier to see in theta_phi than phi.

\begin{figure}
\centering
\includegraphics[width=\hsize]{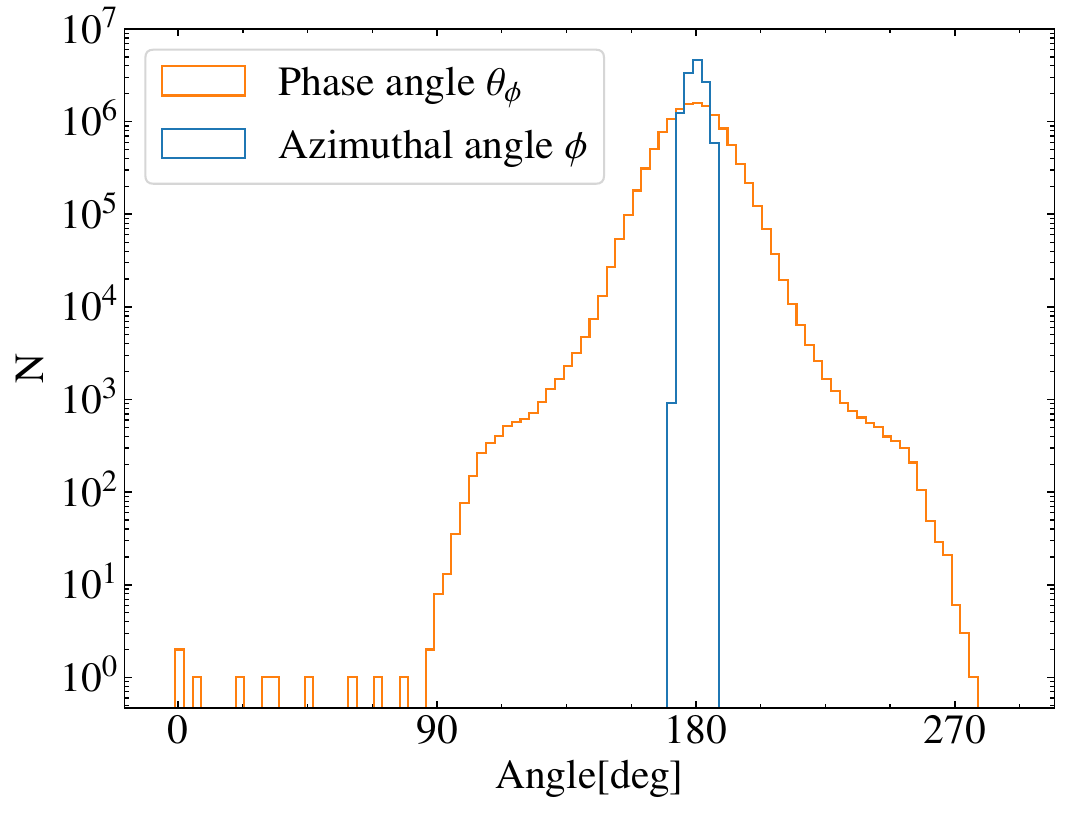}
  \caption{Comparison of the distributions of azimuthal angle $\phi$ and azimuthal angle of guiding centre $\theta_\phi$ for stars in the solar neighbourhood ($d < 1$ kpc).
          }
     \label{fig:theta_v_theta_phi_distribution}
\end{figure}

Following \cite{hunt_resolving_2021}, who showed that dividing a sample of stars by the azimuthal angle of their guiding centres (also called the phase angle, $\theta_\phi$) reveals phase spiral patterns more clearly than dividing by the azimuthal angles of the stars themselves ($\phi$), we compute action-angle variables for our sample.
We use the galactic dynamics package \verb|AGAMA| \citep[Action-based Galaxy Modelling Architecture,][]{vasiliev_agama_2019} with the \verb|best| potential from \cite{mcmillan_mass_2017}.

A Gaussian filtering technique was used to enhance the contrast of the phase spiral in our plots. We utilised the \verb|scipy.ndimage.gaussian_filter| function from the \verb|SciPy| library \citep{virtanen_scipy_2020} to create a blurred version with the Gaussian filter and dividing the original with it. In this way, the spiral pattern is enhanced.

Figure \ref{fig:nearby_theta} compares four plots of the phase spirals in the solar neighbourhood to demonstrate the effect that a Gaussian filter and the choice of angle variable make.
All of those plots use stars within 1\,kpc of the Sun and with angular momentum in the $1400 < L_Z \,/ \,\mathrm{kpc\, km\, s}^{-1} < 1500$ range, meaning they have a guiding centre distance close to 6.3\,kpc. This is within the range of $R_G$ where \cite{hunt_multiple_2022} discovered the two-armed phase spiral.
The left column has stars selected by their azimuthal angle, in the range $175^\circ < \phi < 185^\circ$ and the right column has stars selected by their phase angle, in the range $175^\circ < \theta_\phi < 185^\circ$. 
When using $\phi$, 469\,592 stars are selected and when using $\theta_\phi$, 186\,447 stars are selected.
The upper plots show number density and the bottom plots show number density contrast, computed with the Gaussian filter.
This filter is scaled to the non-square pixels.
The kernel has a scale of 500\,pc in the $Z$ direction and 40\,km\,s$^{-1}$ in the $V_Z$ direction.
It is evident that the appearance of the two-armed spiral pattern is much clearer when using number density contrast and selecting by $\theta_\phi$ compared with selecting by $\phi$, despite the selection by $\phi$ containing many more stars.

Figure \ref{fig:theta_v_theta_phi_distribution} shows a comparison of the distributions of azimuthal angle $\phi$ and phase angle $\theta_\phi$ for stars in the solar neighbourhood ($d < 1$ kpc).
The $\phi$ angles are naturally limited to being close to $180^\circ$ while $\theta_\phi$ has a significantly larger spread, ranging from $90^\circ$ to $270^\circ$, demonstrating that we are sampling a large part of the Galaxy even when we select stars close to the Sun.

\subsection{Location of the two-armed phase spiral}
% The two-armed spiral is only detected in solar neighbourhood and only at low L_Z.
% The two-armed spiral can be seen in number density, radial velocity, and (faintly) in mean metalicity

\begin{figure}
\centering
\includegraphics[width=\hsize]{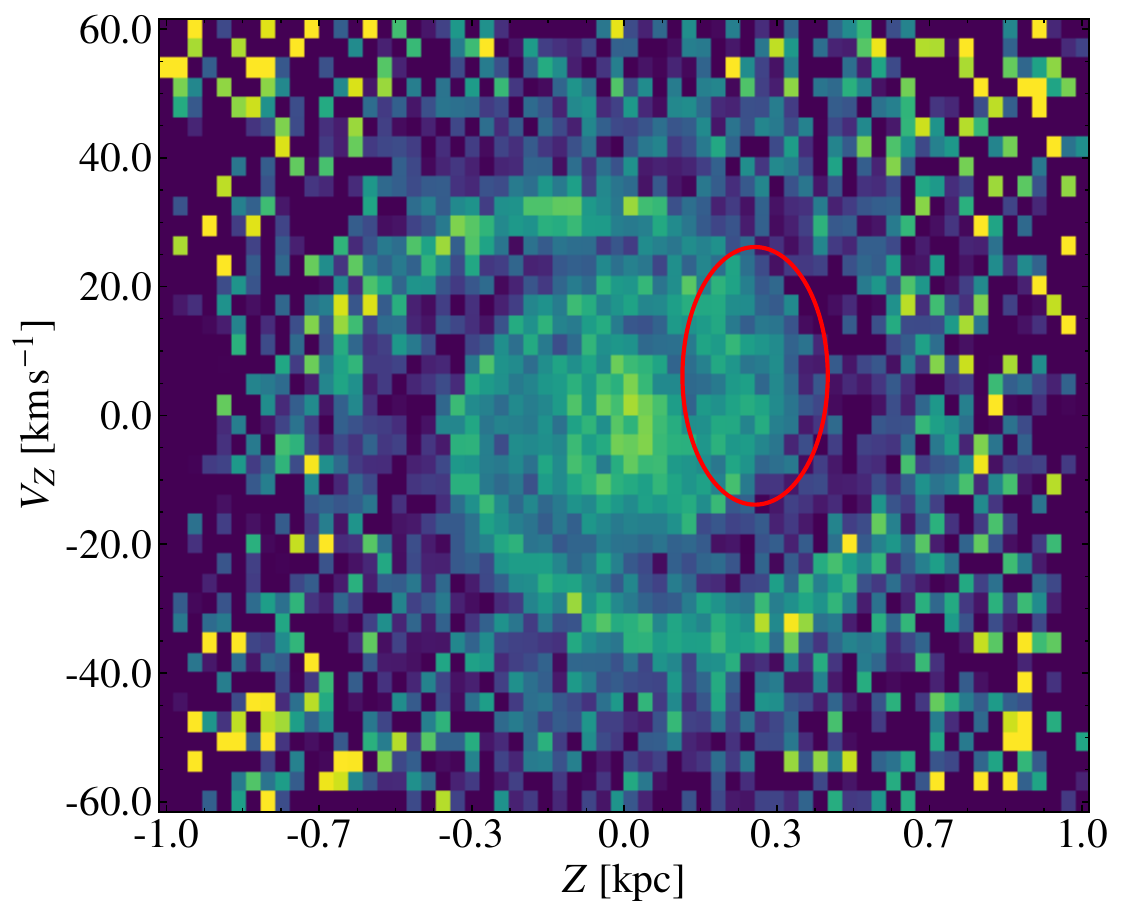}
  \caption{The two-armed phase spiral in the region 
    $1400 < L_Z \,/ \,\mathrm{kpc\, km\, s}^{-1} < 1500$,
    $ d < 1 \,\mathrm{kpc}$,
    $175^\circ < \theta_\phi < 185^\circ$.
    The area that both spiral arms appear to emerge from is marked with a red ellipse.
          }
     \label{fig:annotated}
\end{figure}

\begin{figure*}
\centering
\includegraphics[width=\hsize]{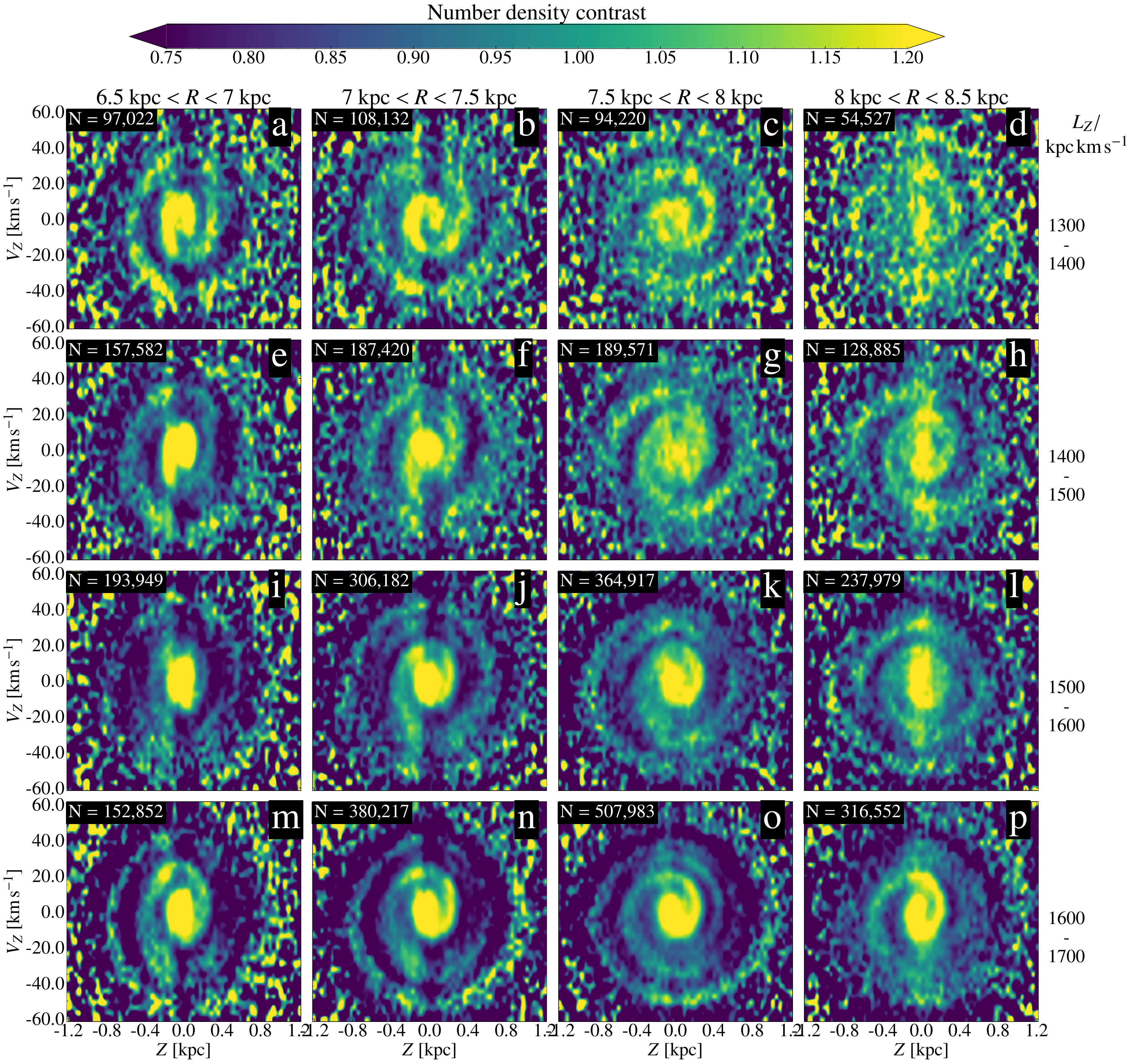}
  \caption{The $Z$-$V_Z$ phase plane at different radii (left to right) and angular momenta (top to bottom), showing the one- or two-armed phase spiral. In all cases, the data is restricted to $170^\circ < \theta_\phi < 190^\circ$. The quantity indicated by the colour bar is the number density contrast after processing with our Gaussian filter.
          }
     \label{fig:grid-two-arms}
\end{figure*}

\begin{figure*}
\centering
\includegraphics[width=\hsize]{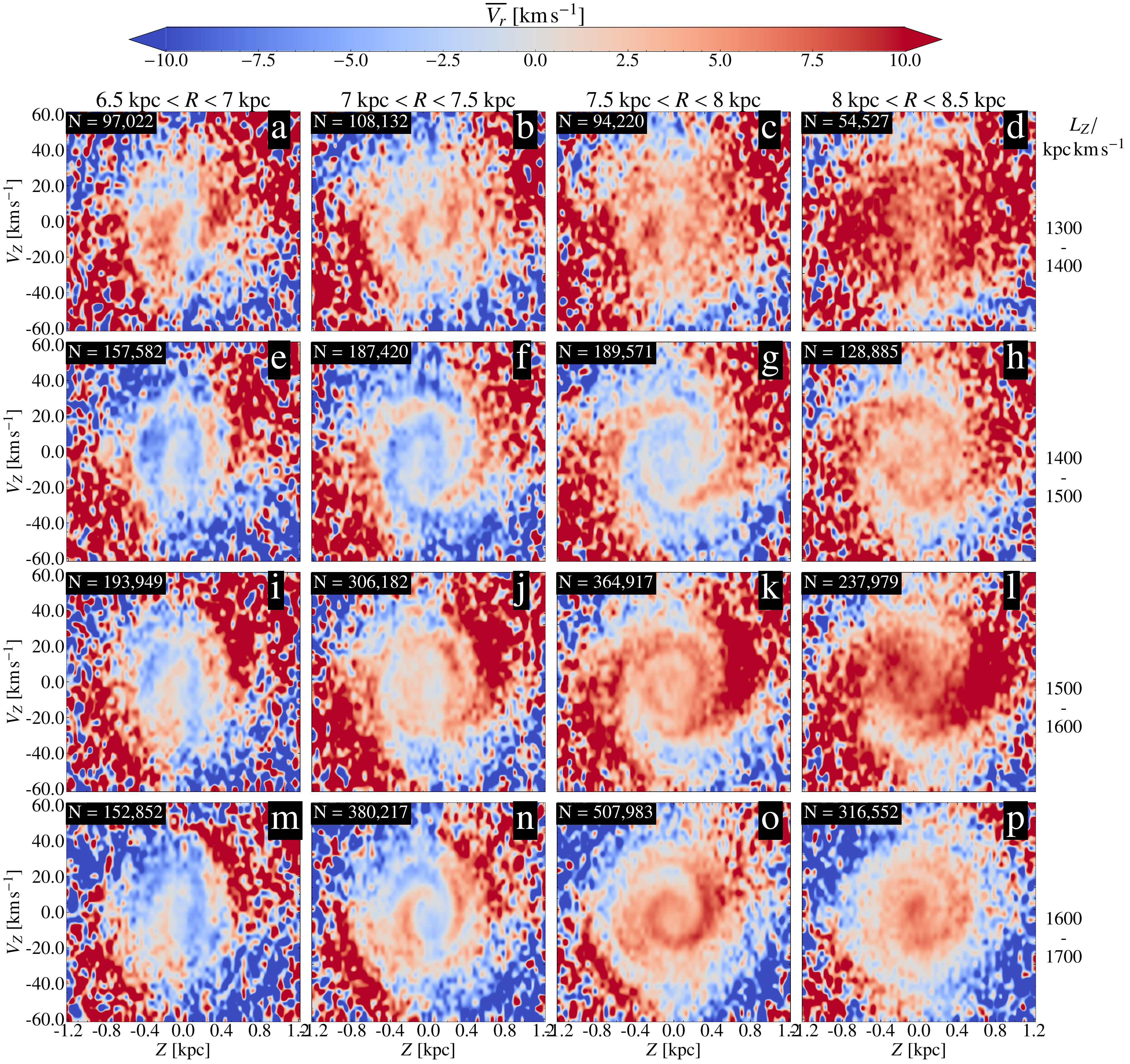}
  \caption{The $Z$-$V_Z$ phase plane coloured by galactocentric radial velocity. Individual plots are at different radii (left to right) and angular momenta (top to bottom). In all cases, the data is restricted to $170^\circ < \theta_\phi < 190^\circ$. %[say something about phase spiral?]
          }
     \label{fig:grid-vr}
\end{figure*}

Figure~\ref{fig:annotated} shows the two-armed phase spiral in number density contrast with a Gaussian filter for stars in the solar neighbourhood with low angular momentum, $L_Z = 1450 \; \pm 50 $ kpc km s$^{-1}$.
This plot establishes some characteristics of the two-armed phase spiral. 
Mainly it appears like the one-armed phase spiral seen at higher angular momentum but with an extra arm emerging from the same ``root'' as the first one (marked with a red ellipse) but winding slower and ending before it would cross into the negative $Z$-range.

Figure~\ref{fig:grid-two-arms} shows the stellar number density contrast in multiple regions of $R$ and $L_Z$. 
We can see that the two-armed phase spiral only appears in a limited region around panel g, with $R \approx 8$\,kpc and $ L_Z \approx 1450$\,kpc\,km\,s$^{-1}$.
We also see a one-armed phase spiral at higher angular momentum in Figs.~\ref{fig:grid-two-arms}b and \ref{fig:grid-two-arms}c, and at the same or lower angular momentum in Figs.~\ref{fig:grid-two-arms}o and \ref{fig:grid-two-arms}p.

In Fig.~\ref{fig:grid-vr}, we see the same sample of stars coloured by mean galactocentric radial velocity. Here we see a two-armed spiral pattern in the middle two rows, between 1400 and 1600\,kpc\,km\,s$^{-1}$.
Since the mean radial velocity is much less affected by dust, this is a useful method for tracing phase spirals \citep{li_gaia_2023}.

\begin{figure*}
\centering
\includegraphics[width=\hsize]{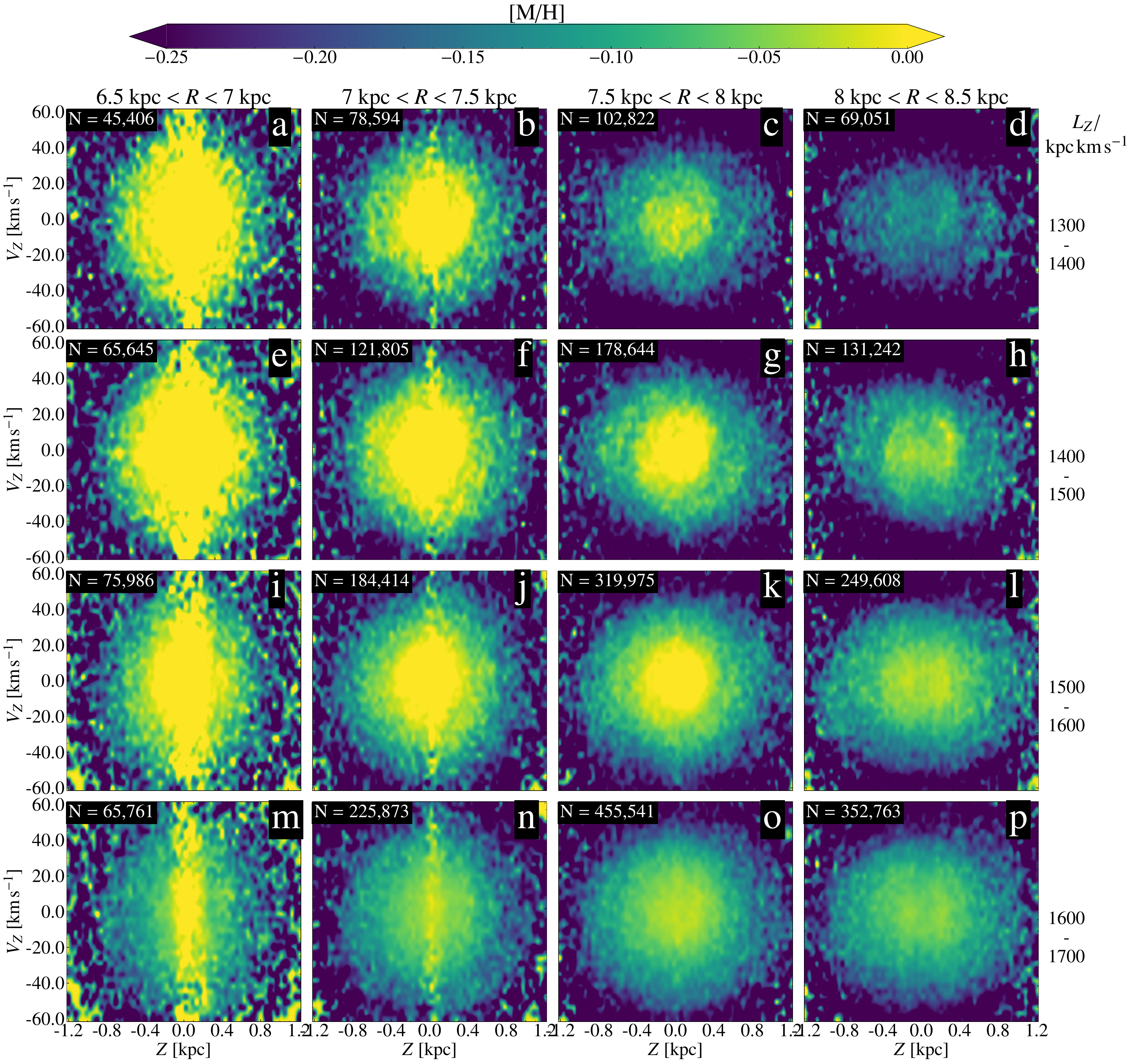}
  \caption{The two-armed phase spiral at different locations, coloured by mean metallicity. In \textit{panel g}, we can see a trace of a spiral arm in the negative $z$, positive $v_z$ quadrant.
          }
     \label{fig:metal_grid}
\end{figure*}

Figure~\ref{fig:metal_grid} shows the same regions as Fig.~\ref{fig:grid-two-arms} but coloured by mean metallicity. 
Here we can see two faint ridges corresponding to a two-armed spiral pattern (Fig.~\ref{fig:metal_grid}g), the same panel that shows the strongest pattern in Fig.~\ref{fig:grid-two-arms}.
This indicates that the stars participating in the two-armed phase spiral are of a slightly higher mean metallicity than others at the same vertical energy, at the same distance from the centre of phase space.
This suggests that these stars have been transported from a region of space or phase space with higher mean metallicity.
Given the well-known vertical and radial gradients in metallicity \citep[e.g.,][]{recio-blanco_gaia_2023}, they may have been scattered from an original location at a lower galactocentric radius or a lower amplitude vertical oscillation (or both) by one of the mechanisms that form phase spirals.

\subsection{The two-armed model} \label{ssection:model}
\sidecaptionvpos{figure}{c}
\begin{SCfigure*}[0.4]
\centering
\includegraphics[width=0.7\textwidth]{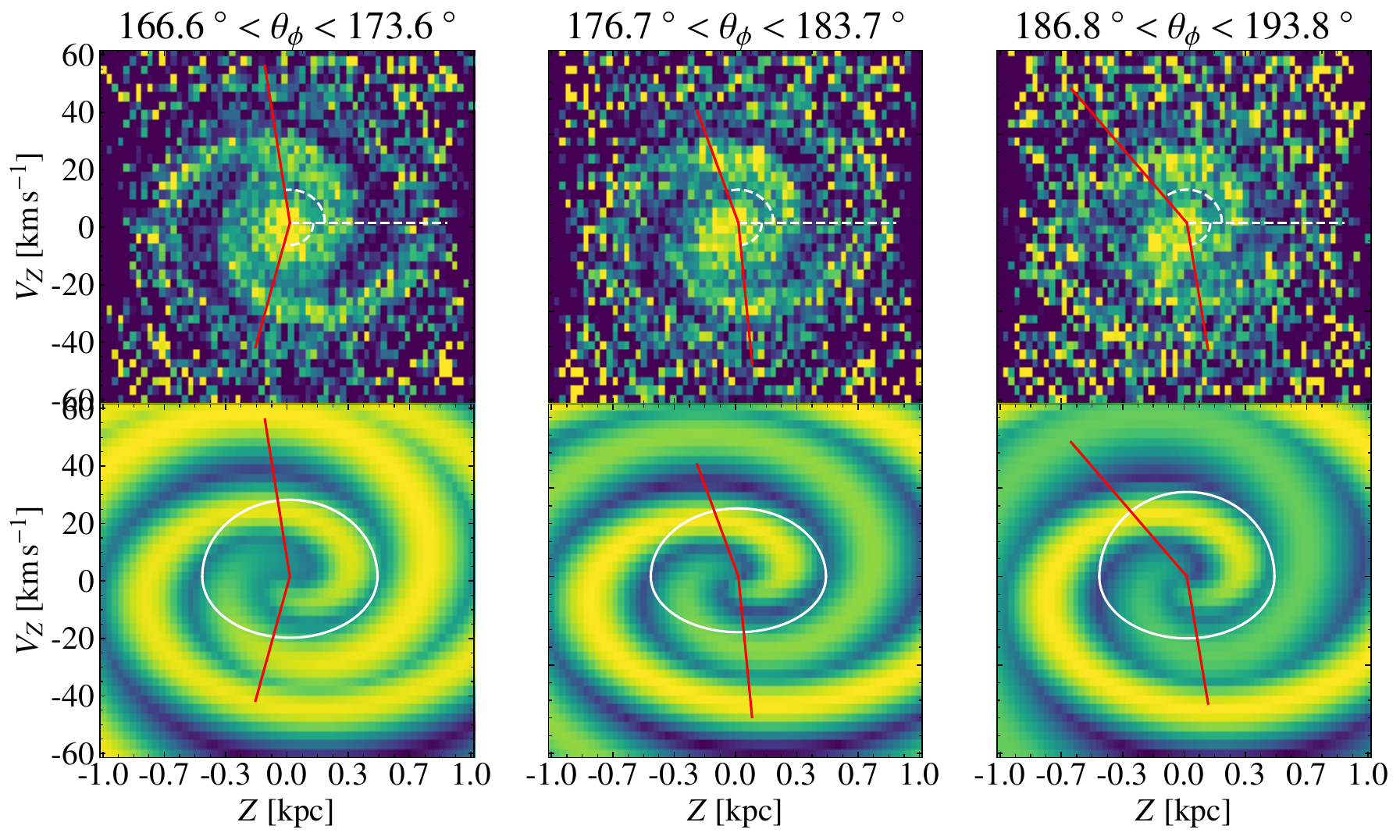}
  \caption{Visualisation of how the angle of the arms of the phase spiral is found.
  \textit{Upper row}: Number density contrast of the two-armed phase spiral at low, medium, and high $\theta_\phi$.
  The spiral angle is marked with a red line and the zero-point for the angle is a dashed white line.
  \textit{Lower row}: Model of the two-armed phase spiral corresponding to the data in the top row.
  The spiral angle and measurement distance are marked with a red line and a white ring respectively.
          }
     \label{fig:angle_compare}
\end{SCfigure*}

\begin{table}
\caption{Free parameters in the model of the phase spiral.}
\centering 
\begin{tabular}{ccccc}
\hline \hline
Name & min & max & Description & Unit \\
\hline
$\alpha$ & 0 & 1 & Amplitude & -- \\
$b$ & 0.005 & 0.1 & Linear winding & $\mathrm{pc\,rad^{-1}}$ \\
$c$ & 0.0 & 0.004 & Quadratic winding & $\mathrm{pc\,rad^{-2}}$ \\
$\theta_0$ & $-\pi$ & $\pi$ & Angle offset & rad \\
$S$ & 30 & 70 & Scale factor & $\mathrm{km\,s^{-1}\,kpc^{-1}}$ \\
$\rho$ & 0.0 & 0.18 & Flattening distance & kpc \\
\hline
\end{tabular}
\label{tab:priors} 
\tablefoot{For each parameter we assume a prior probability that is constant between these min and max values.}
\end{table}

We use a model to quantify the properties of the two-armed phase spiral.
The model uses a Markov Chain Monte Carlo approach to fit a mathematical spiral defined by
\begin{equation}
	r = b\phi_s + c\phi_s^2
\label{eqn:main}
\end{equation}
to a 2D histogram of an observed phase spiral in phase space.
The model creates a smooth background of the phase space distribution of stars which removes all the detailed features of the distribution, including the arms of the phase spiral.
The model then tries to find a set of parameters for a mathematical spiral such that when it is multiplied with the smooth background, it recreates the original distribution.
The model uses the scale factor $S$ to measure the axis ratio of the phase spiral.
It works by stretching the $V_Z$ axis such that a higher scale factor increases the vertical velocity extent of the phase spiral.
This is needed as the phase spiral is known to change shape, especially with galactocentric radius \citep[e.g.,][]{laporte_footprints_2019,li_vertical_2021}, and the scale factor accounts for this in the model.
This is done by changing the way distances and angles are computed in the following equations,
\begin{equation}
	r = r(Z, V_Z) = \sqrt{Z^2 + \left( \frac{V_Z}{S} \right)^2 },
\label{eqn:r}
\end{equation}
\begin{equation}
	\theta = \theta (Z, V_Z) = \arctan \left( \frac{1}{S} \frac{V_Z}{Z} \right).
\label{eqn:theta}
\end{equation}

For the full equation for the phase spiral, we take inspiration from \cite{widmark_weighing_2021-1} and end up with
\begin{equation}
	f(r, \theta) = 1 + \alpha \cdot \mathrm{sigm} \left( \frac{r-\rho}{0.1 \, \mathrm{kpc}} \right) \cos(\theta - \phi_s(r)-\theta_0 ).
\label{eqn:full}
\end{equation}
All free parameters are named in Table~\ref{tab:priors} and $\phi_s, r, $ and $\theta$ are defined in Eqs.~\ref{eqn:main}, \ref{eqn:r}, and \ref{eqn:theta} respectively.
For a full and detailed description of the model and all parameters, see \cite{alinder_investigating_2023} where it was initially developed.

The model used in this paper is modified to consist of two independent phase spirals combined together after we applied a Gaussian filter to the phase plane.
We can extract the parameters of each of the two phase spirals separately.
When the modified model compares the mathematical spiral to the phase space distribution, instead of making the comparison right away, it creates two spirals and combines them using an elementwise maximum function.
Rather than attempting to fit these spirals to the normal number density distribution, the modified model attempts to match the spirals to the distribution after it has been processed with a Gaussian filter.
This filter uses a kernel size that corresponds to 500\,pc in the $Z$ direction and 40\,km\,s\,$^{-1}$ in the $V_Z$ direction, and has had noise suppressed by clipping pixels with values smaller than the 5th percentile or larger than the 95th percentile.

We also want to be able to measure the rotation of the phase spiral with this model.
The model parameter $\theta_0$ represents the angle offset, which is the rotation of the phase spiral in the model.
However, this parameter is not a convenient descriptor of the phase spiral's rotation because it has a degeneracy with the linear winding parameter $b$, and also to a certain extent with the quadratic winding parameter $c$.
Different sets of these values can produce very similar spirals except in the most central regions, which are softened through the flattening parameter.
Therefore, we describe the rotation of the phase spiral by measuring the angle of the spiral arm at a fixed phase distance.
These angles are shown in Fig.~\ref{fig:angle_compare} with red lines, and the phase distance the measurements are made at is indicated with a white ring (scaled to the same axis ratio as the phase spirals) in the lower row. 
The angle $0^\circ$ is shown with a dashed white line in the upper row.
Changing the phase distance at which we take these measurements does not change our results significantly except by changing all angles by a constant amount.

The model used in this paper has slightly different allowed parameter ranges than \cite{alinder_investigating_2023}, and these are listed in Table \ref{tab:priors}.
These changes allow the model to find a tighter phase spiral by reducing the two winding parameters and make it pay more attention to the inner part of the phase spiral by reducing the maximum flattening distance.
The winding parameters specify how tightly the phase spiral turns with radial distance.
The flattening distance $\rho$ is the size of an area in the centre of the phase plane where the spiral function is ``flattened'' to reduce the significance of the central region since it is not fitted well by this kind of model.
If these changes are not made, the model may explore solutions with very tight spirals (large values of the winding parameters in Eq.~\ref{eqn:main}) and get stuck in false minima with unreasonable solutions.
The changes are:
\begin{itemize}
	\item $b$ changed to the range $0.005 - 0.1$ from $0.01 - 0.175$,
	\item max $c$ reduced to 0.004 from 0.005,
	\item Max flattening distance reduced to 0.18 from 0.3.
\end{itemize}

\subsection{Rotation}
% Rotation of the two-armed spiral is observed
% This rotation can be quantified with the model
% Briefly explain how the model works and what was changed
% The one-armed spiral also rotates in theta_phi
% The rotation is also visible in V_r
% With the model we can detect a small differance in the rate of rotation of the arms (confirm)
% If possible, show the different way the two-arms rotate

\begin{figure}
	\centering
	\includegraphics[width=\hsize]{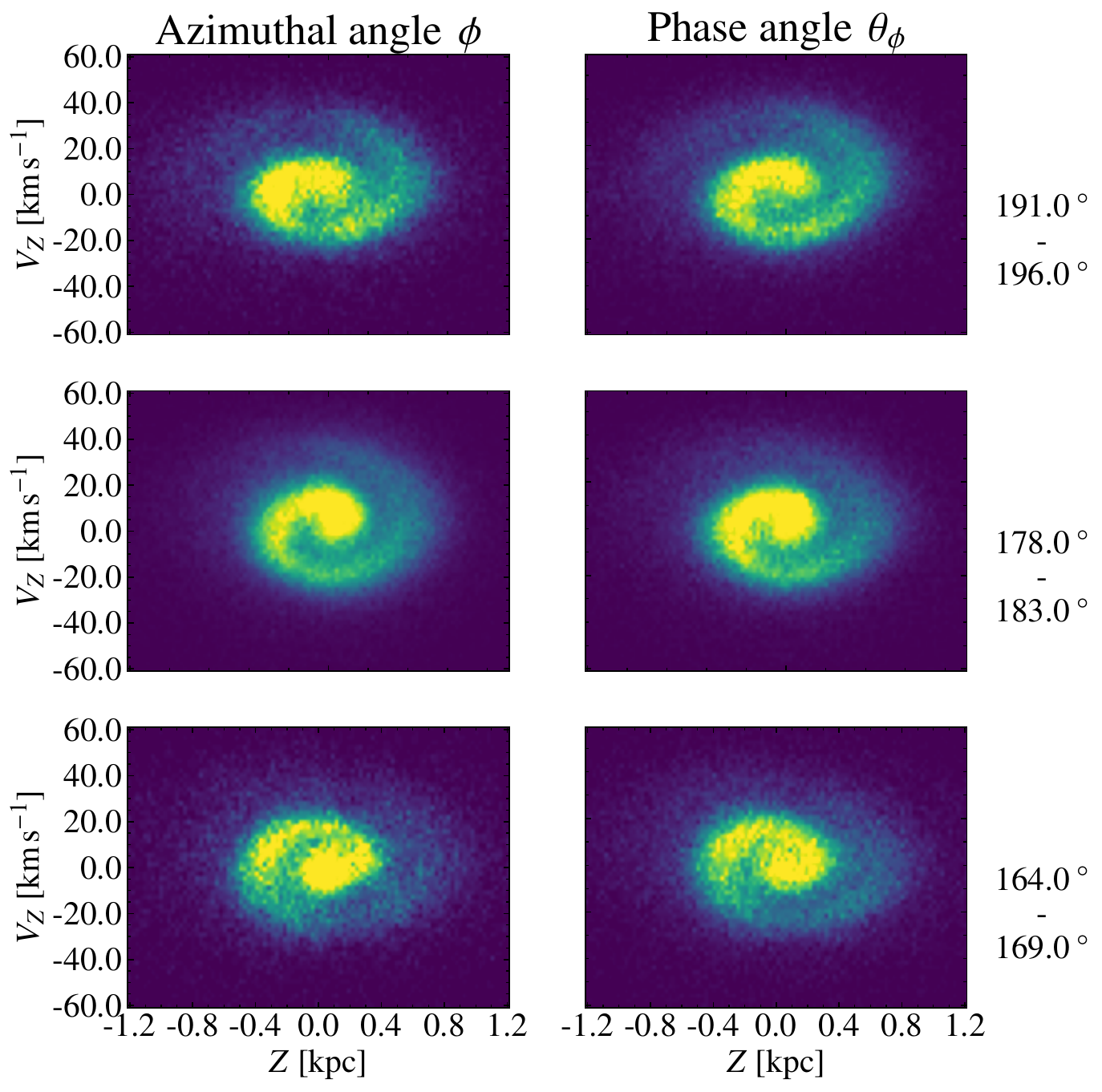}
	\caption{Comparison of the normalised number density phase spiral in the outer Galactic disc sliced by azimuthal angle and phase angle.
    An animated version is available at \href{https://lu.box.com/s/7qgvkg4jp5jdnm4m4cnq5smnkux018gt}{this link}.
	}
	\label{fig:anim_snapshot1}
\end{figure}

\begin{figure*}
\centering
\includegraphics[width=\hsize]{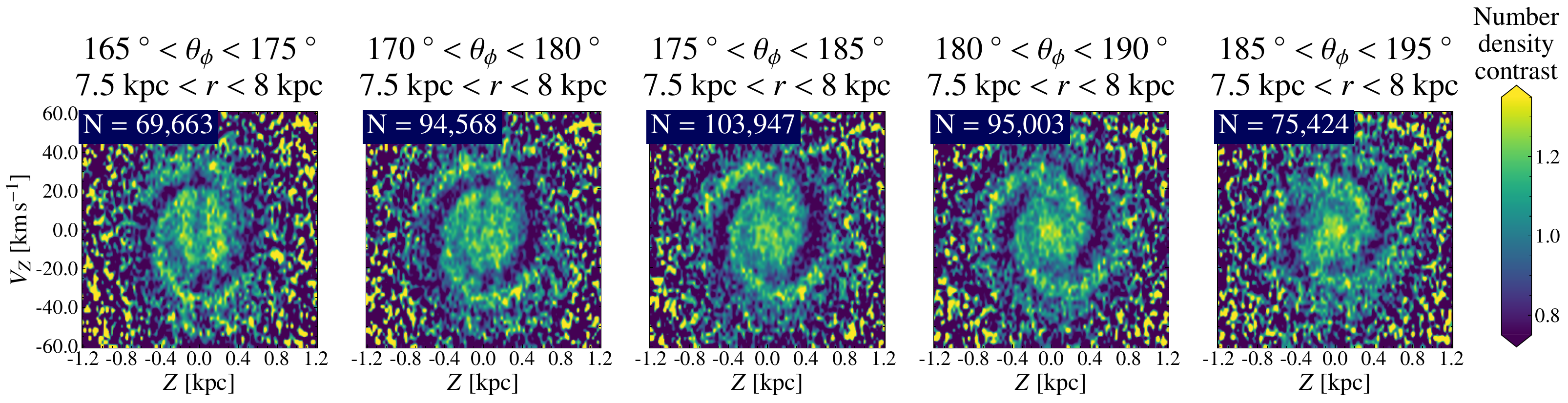}
  \caption{The number density contrast in the phase plane showing the two-armed phase spiral at different values of $\theta_\phi$.
  The stars are in the angular momentum range of 1400 to 1500 $\mathrm{kpc\, km\, s}^{-1}$ and at a galactocentric distance of 7.5\,kpc to 8\,kpc.
  The phase spiral appears to rotate counter-clockwise with increasing phase angle.
  The data presented in this figure can also be seen as an animation at \href{https://lu.box.com/s/x8zmq29pcyubcmeqyraqcuaa5oqov7s3}{this link}.
          }
     \label{fig:rotate}
\end{figure*}

\begin{figure}
	\centering
	\includegraphics[width=\hsize]{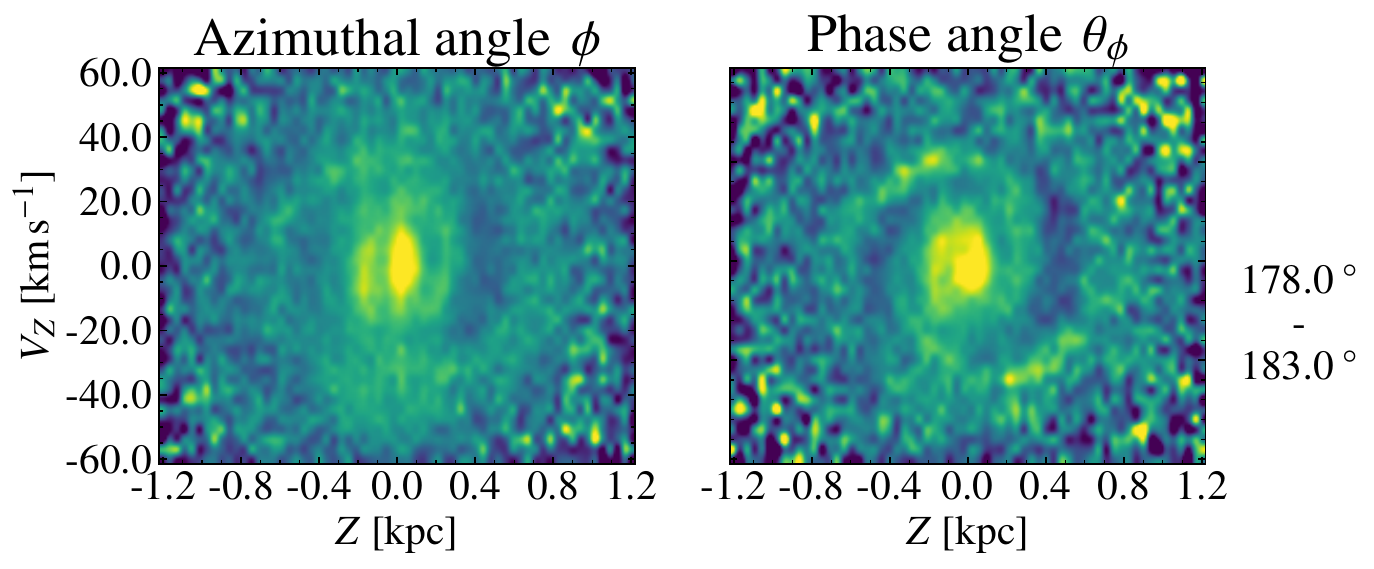}
	\caption{Comparison of the two-armed phase spiral with stars selected using azimuthal angle $\phi$, \textit{left}, and phase angle $\theta_\phi$, \textit{right}, both in the range from $178^\circ$ to $183^\circ$.
    The\textit{ right panel} shows a relatively clear two-armed phase spiral, while it is relatively faint in the \textit{left panel}.
    An animated version of this figure is available at \href{https://lu.box.com/s/qxoguxfeiivsoiof4zt2gt712xancqhj}{this link}.
	}
	\label{fig:anim_snapshot2}
\end{figure}

\begin{figure}
\centering
\includegraphics[width=\hsize]{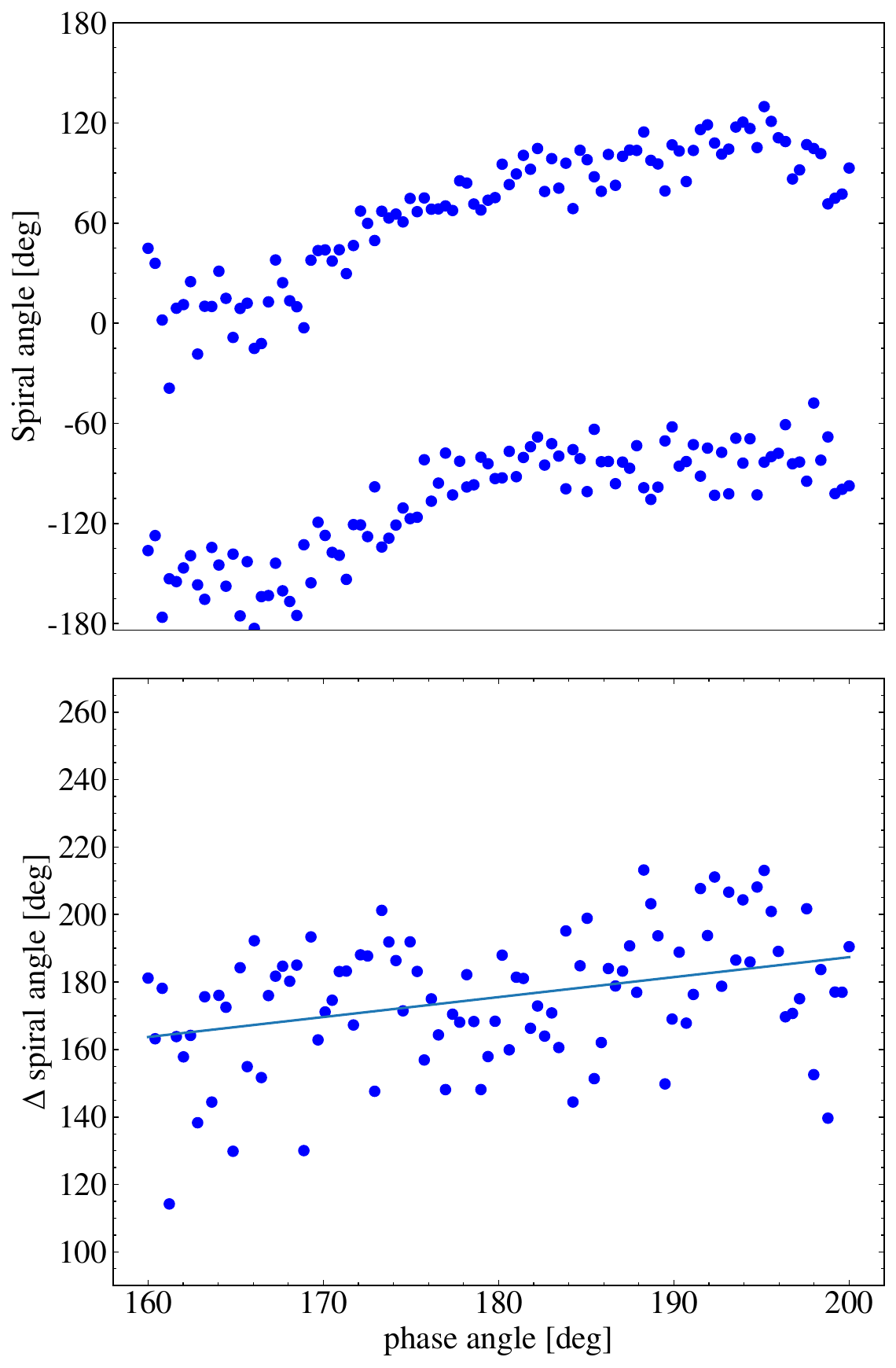}
  \caption{\textit{Top panel}: The rotation angle of the phase spirals as functions of azimuthal angle of guiding centre $\theta_\phi$ for stars in the solar neighbourhood ($d < 1$\,kpc).
  Each point uses data from $\pm 3.5^\circ$.
  The zero point of the spiral angle is arbitrary.
  \textit{Bottom panel}: The difference in the angle of the arms of the phase spiral.
  The trend is fitted with a line.
          }
     \label{fig:dual_theta_trend}
\end{figure}

% Recap the behaviour of the phase spiral in the outer disc from paper 1.
In \cite{alinder_investigating_2023}, we investigated how the phase spiral changes with azimuthal angle and found that the phase spiral in the outer Galactic disc rotates with azimuthal angle $\phi$ at a rate of about $3^\circ$ per degree of $\phi$.
This region is an angular momentum range of $2200 < L_Z / \mathrm{kpc\, km\, s}^{-1} < 2400$ and a galactocentric radial range of $8.5 < R / \mathrm{kpc} < 10.5$. 
In this study, we use the phase angle $\theta_\phi$, instead of the azimuthal angle $\phi$, when selecting stars.
To ensure the results of this paper are comparable with those of \cite{alinder_investigating_2023}, we plot the phase spiral in the outer Galactic disc using both $\phi$ and $\theta_\phi$.
Figure \ref{fig:anim_snapshot1} shows the phase spiral in the outer Galactic disc at different values of $\phi$ (left column) and $\theta_\phi$ (right column). An animated version of this comparison can be seen \href{https://lu.box.com/s/7qgvkg4jp5jdnm4m4cnq5smnkux018gt}{here}\footnote{The animation covers $210^\circ$ to $150^\circ$. The frames have large overlaps with each frame covering $5^\circ$.}.
The figure shows a comparison of the phase spiral plotted using stars in three different ranges of phase angle $\theta_\phi$ or azimuthal angle $\phi$ values.
The general appearance and rotational behaviour of the phase spiral are similar in both columns.
Both columns of the figure show the clearly identifiable phase spiral in the outer Galactic disc rotating approximately half a rotation over the shown range of angles.
The phase spiral plotted using $\phi$ appears to be more strongly affected by dust at high and low angles as there is a noticeable decrease in the number density of stars around $Z = 0$\,pc.
This effect is present but much weaker in the phase spiral drawn using  $\theta_\phi$, as not all stars in the frames with high or low angles are physically distant from the Sun.

% Relate the properties of the outer spiral to the two-armed spiral. \ref{fig:anim_snapshot2}
We want to compare the rotation of the one-armed phase spiral in the outer Galactic disc, seen in Fig.~\ref{fig:anim_snapshot1}, with the rotation of the two-armed phase spiral as seen in the angular momentum range of $1400 < L_Z / \mathrm{kpc\, km\, s}^{-1} < 1500$ in the solar neighbourhood, as illustrated in Fig.~\ref{fig:rotate}.
In order to do so we plot the number density contrast of the phase plane of stars in this angular momentum range and within 1\,kpc of the Sun in a $5^\circ$ slice around $180^\circ$ in both $\phi$ and $\theta_\phi$, shown in Fig.~\ref{fig:anim_snapshot2}\footnote{This comparison is also available as an animation \href{https://lu.box.com/s/qxoguxfeiivsoiof4zt2gt712xancqhj}{here}.
The animation shows the number density contrast from $\phi$ and $\theta_\phi$ in the range $200^\circ$ to $160^\circ$ with each frame covering $5^\circ$.
The rotation seen in $\theta_\phi$ is approximately $90^\circ$ over this range.
In  $\phi$, a phase spiral can only barely be seen in the densest part, around $\phi = 180^\circ$, but a rotation is hard to discern.}.
The left panel is plotted using $\theta_\phi$.
In it, both arms of the phase spiral can be seen while in the right panel, plotted using $\phi$, the spiral pattern is much less clear but still vaguely detectable.
Since the phase spiral in this region rotates with $\theta_\phi$, and the phase spiral in the outer Galactic disc rotates with both $\phi$ and $\theta_\phi$, it is reasonable to assume the reason for this rotation is similar or the same in the two regions.

% Discuss the details of the two-armed spiral.
In Fig.~\ref{fig:rotate} we can see how the phase spiral changes with $\theta_\phi$ angle.
The figure shows the number density contrast with a range of $\theta_\phi$ going from $195^\circ$ to $165^\circ$ with each panel covering $10^\circ$. 
Both arms of the phase spiral can be seen to rotate and the range of rotation is approximately $90^\circ$ over this range of $\theta_\phi$. 
The phase spiral rotates clockwise with decreasing $\theta_\phi$, like the phase spirals observed in \cite{alinder_investigating_2023}. 
% Conslusion? Or is it fine like this?

Now we employ our model, discussed in Sect.~\ref{ssection:model}, to fit the two-armed phase spiral.
We want to measure how the phase angle $\theta_\phi$ affects the rotation of the phase spiral and we want to measure the difference in angle between the two spirals over a range of phase angles.
In Fig.~\ref{fig:dual_theta_trend}, we can see the results of the model. 
In the upper panel, we see two tracks.
These two tracks are the orientation of the two arms of the phase spiral, called Spiral angle in the figure, as identified by our model.
Over the range of $\theta_\phi = 160^\circ$ to $\theta_\phi = 200^\circ$, the arms of the phase spiral move from a spiral angle of approximately $0^\circ$ to $90^\circ$ and $-180^\circ$ to $-90^\circ$ respectively.
The bottom panel shows the difference in spiral angle between the two arms.
The distance stays mostly consistent but a least-squares fitted line reveals that the separation is increasing at a rate of about $0.5^\circ$ in $\Delta$ spiral angle per $1^\circ$ in $\theta_\phi$ in this range.

\subsection{Axis ratio}
\begin{figure}
	\centering
	\includegraphics[width=\hsize]{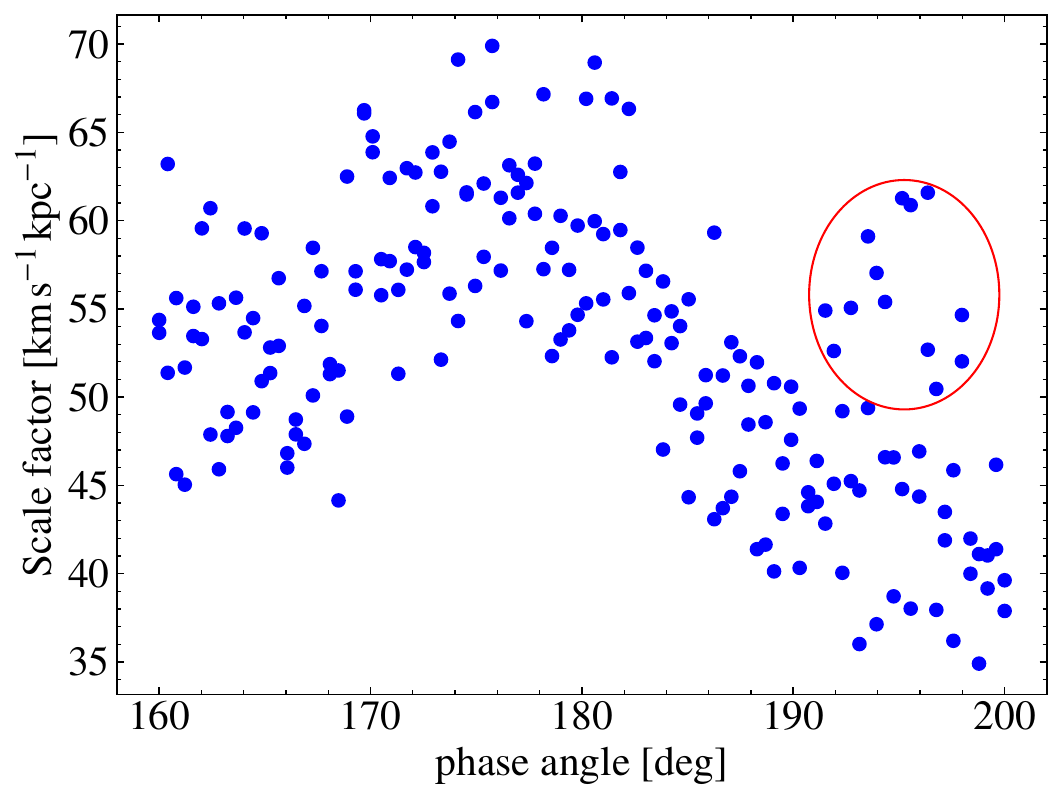}
	\caption{Scale factor of the model as a function of phase angle $\theta_\phi$. The scale factor is equivalent to the axis ratio of the phase spiral.
    The scale factor shows a peak at around $\theta_\phi = 175^\circ$.
    The points contained by the red ellipse are off the apparent sequence, and it is unclear whether to attribute them to difficulties fitting the phase spirals in that region or a genuine increase in scale factor.
	}
	\label{fig:scale_factor}
\end{figure}

In Fig.~\ref{fig:rotate}, the phase spiral also appears to change shape, becoming less extended in the $Z$-direction and more extended in the $V_Z$-direction.
The axis ratio is parametrised in the model with the scale factor $S$, so it can be examined in the same way as the rotation angle above.
Figure \ref{fig:scale_factor} shows the values for the scale factor for the same model run seen in Fig.~\ref{fig:dual_theta_trend}.
The figure shows a distribution with a large spread but a clear peak at around $\theta_\phi = 175^\circ$.
It is unclear to us whether the $\sim 10$ points marked by the red ellipse at $\theta_\phi > 190^\circ$ above the main line are spurious, due to large uncertainties, or indicate a genuine increase in scale factor in this region.

This dependence of the axis ratio on a spacial location has been seen previously by, for instance, \cite{widmark_weighing_2022-1} and \cite{guo_measuring_2024} and is usually taken to be a probe of the vertical potential of the disc.
Assuming this is the case, we encourage future studies into how to use the varying shape of the two-armed phase spiral to investigate azimuthal differences in the potential of the galactic disc.

\section{Discussion}\label{section:discussion}
% Speculate about if the two-armed spiral has a different origin than the one-armed one
% Speculate about what mechanisms might cause the observed differences, or at least point out if they can't be the same
% Talk about the complexity of the Galaxy and the disc

Vertical disturbances of the Milky Way's disc have been a subject of research for several decades.
The warp of the Galactic disc has been recognised since \cite{kerr_magellanic_1957} found it in the gas component and \cite{freudenreich_dirbe_1994} in the stellar component, showing that the Galactic disc is not a simple, flat structure.

With the data made available by the \textit{Gaia} mission, a much fuller view of the complexities of the structures in the Galactic disc has been revealed, with numerous studies investigating different views of the Galactic disc's disturbances.
\cite{poggio_galactic_2018} mapped the warp in populations of both young and old stars out to 7\,kpc from the Sun.
\cite{antoja_dynamically_2018} found the signal of the warp as a vertical velocity gradient in both populations in the outer Galactic disc and claimed this supports the hypothesis that the warp is a gravitationally induced phenomenon.
\cite{ramos_riding_2018} found numerous ridges in $V_\phi - R$ space and suggest the various structures they see may be the results of several different mechanisms acting on the disc.
\cite{gaia_collaboration_gaia_2021-1} investigated the velocity distribution of stars in the outer Galactic disc and found it asymmetric with respect to the plane of the Galaxy.
They also found the $V_Z - V_\phi$ distribution to be bimodal, forming two clumps, as well as additional ridges in $V_\phi - R$ space.
%\cite{gaia_collaboration_gaia_2023-1} talked about asymmetries in the disc.
\cite{hunt_radial_2024} found traces of phase spirals in $R - V_R$ space in the solar neighbourhood which may have a different origin than the $Z - V_Z$ phase spirals discussed in this paper.
\cite{cao_radial_2024} looked at waves in the $L_Z - \langle V_R \rangle$ space which they suggested have an origin that is internal to the Galaxy, and may possibly be useful for inferring information about the properties of the bar.
It seems likely that many of these phenomena have the same root cause, and that what we are really observing are different facets of a grand and complicated dynamical system.

It is reasonable to assume that the properties of the two-armed phase spiral can, in some way, be related to its formation.
It is unclear why the phase spiral has two arms in just the specific range found in this investigation and not elsewhere.
It might be because the formation mechanism results in a weak two-armed pattern, that the two-armed phase spiral used to exist across a large range but has been attenuated over time, or that the mechanism works strongest in a different region of the Galaxy and we are only detecting the edges of the stronger pattern produced by that mechanism.
The last option seems unlikely as it would imply that it should be possible to detect a stronger signal by looking in a different, presumably adjacent, region, which has been attempted in this paper without success in Sect.~\ref{section:results}.

\cite{hunt_multiple_2022} argue that the two-armed phase spiral is a response to a ``breathing mode'', that is, compression and rarefaction of stars vertically in the Galactic disc. This response has been found in theoretical calculations by \cite{banik_comprehensive_2022} and in $N$-body simulations by \cite{hunt_resolving_2021}.
As \cite{hunt_multiple_2022} note, the most likely way to produce this in the inner part of the Galactic disc is by internal perturbations, rather than the response to an interaction with a satellite galaxy.
If this is indeed the case, possibly due to strong spiral arms within the disc, it would suggest the two-armed phase spiral is a rather localised effect.
If the two-armed phase spiral indeed has a different origin from the one-armed phase spiral, that might account for some of the differences in observed properties.

% If Li has a paper/preprint out talking about the two-armed phase spiral as possibly caused by two separate impacts, put a paragraph here about it.
% From my memory of whatwe said at the conference, a large part of the work was using the difference in phase angle to estimate the time between the impacts.

\cite{widmark_weighing_2021-1} uses the structure and shape of the one-armed phase spiral to infer the gravitational potential of the Milky Way disc.
This intuitively fits with the observed increase in vertical velocity range of the phase spiral with increased galactocentric radial distance \citep[e.g.,][]{li_vertical_2021}.
The change in axis ratio with phase angle illustrated in Figs.~\ref{fig:rotate} and \ref{fig:scale_factor} would therefore appear to indicate that there is also a change in Galactic disc potential with azimuthal angle (corresponding to phase angle).
This can happen due to the spiral arms, but it would certainly appear that the variation with phase angle that we see is too large to be simply explained by this.
A crude estimate of the relation of our axis ratio to the gravitational potential would note that this axis ratio is likely to be proportional to the vertical orbital frequencies, such as the vertical epicycle frequency, and that these are proportional to the square root of the potential gradient, and therefore roughly proportional to the square root of the surface density.
The variation in axis ratio we observe would therefore imply surface density variations of at least an order of $50$\,percent, which are well beyond those that we believe occur due to the Milky Way's spiral arms. %\cite{}?
It would appear that efforts to constrain the Milky Way's surface density using the two-armed phase spiral will need a more careful understanding of their formation mechanism and propagation through the Galactic disc.

%\cite{li_gaia_2023} claims a non-axisymmetric bar with a changing pattern speed can form a phase spiral, and is required for one to form.
%\cite{hunt_resolving_2021} claim that the outer disc is more likely to retain multiple phase spirals (if they exist). 
%If each passage of an external perturber, like the Sagittarius dwarf galaxy, creates a vertical phase spiral, and the outer disc retains kinematic structures for longer, then it is possible there exists multiple phase spirals in the disc. 
%[It is not clear to me how one can tell the difference between spirals created by interactions with the Sagittarius dwarf galaxy and internal mechanisms in the Milky Way, so I don't fully know what to say about this.]

The result that the two-armed phase spiral is visible in mean metallicity data is consistent with the basic picture of phase spiral production.
A phase spiral is believed to be created when stars in the Galactic disc are acted on by a force, creating an initially synchronised vertical motion.
The stars that have been so displaced will have angular frequencies depending on the extent of their vertical motion such that those with the smallest excursions will oscillate about the midplane with the highest frequency.
This naturally creates a spiral pattern when viewed in $Z - V_Z$ space.
The stars that are part of the phase spiral have therefore % , according to this theory,
been moved out of the midplane of the Galactic disc and the central region of phase space to live in the outer parts of the phase plane.
The stars that otherwise exist in this part of phase space are more likely to be those belonging to the Galactic thick disc and are therefore expected to have lower mean metallicity \citep{bensby_exploring_2014}.
%The stars in the solar neighbourhood on these low angular momentum orbits are expected to show lower mean metallicity than the stars with higher angular momentum since they are from the inner regions of the Galaxy which have lower metallicity on average.

\section{Summary and Conclusions}\label{section:conclusions}
% Summarize data and methods
In this study, we have used astrometric and line-of-sight velocity data from \textit{Gaia} DR3 and stellar metallicities derived by \cite{andrae_robust_2023} from \textit{Gaia} XP spectra to investigate and characterise the properties of the two-armed phase spiral in the solar neighbourhood.
We computed action-angle variables for these stars and used the azimuthal angle of the guiding centre or phase angle ($\theta_\phi$) when selecting stars, rather than the azimuthal position of the star ($\phi$).
This revealed clearer spiral patterns in phase space. A Gaussian filtering technique further enhanced the contrast of the phase spiral in the plots.

% Model
The properties of the two-armed phase spiral were quantified using a model that fits a mathematical spiral to the observed phase space distribution of stars.
The model incorporates parameters capturing the different aspects of the shape of the phase spiral including one, $S$, to account for the changes in axis-ratio.
This model was a version of the model we developed in \cite{alinder_investigating_2023}, modified to fit two spiral arms rather than one.

We found that it is only in a narrow range of angular momentum ($\approx 1400 - 1500$\,kpc\,km\,s$^{-1}$) in the solar neighbourhood, that the phase spiral has two arms.
At both higher and lower angular momentum, the phase spiral has one arm.
This two-armed phase spiral is a faint structure that is easiest to detect in this narrow range of low angular momentum and with a phase angle $\theta_\phi$ near $180^\circ$.
%We, and previous studies like \cite{hunt_multiple_2022}, use a number density contrast with a Gaussian kernel to enhance the detection of the structure.

We were able to show that the two-armed phase spiral rotates with phase-angle around the Galaxy, changing by about $2.25^\circ$ per degree of $\theta_\phi$.
This is similar behaviour to the one-armed phase spiral in the outer Galactic disc which rotates as described in \cite{alinder_investigating_2023} with either azimuthal angle $\phi$ or phase angle $\theta_\phi$.
%The rate of rotation of the two-armed phase spiral with $\theta_\phi$ is about $2.25^\circ$ per degree of $\theta_\phi$.
The difference in angle of the two arms of the two-armed phase spiral appears to increase by about $0.5^\circ$ per degree of $\theta_\phi$.
We were also able to see that the axis ratio of the two-armed phase spiral varies with $\theta_\phi$ in a similar manner to how the axis ratio of one-armed phase spirals varies with galactocentric distance. % Would like a citation here.

We show that the stars in the overdense regions of phase-space associated with the two-armed phase spiral have a slightly higher mean metallicity than those with equivalent vertical oscillations. This is likely to be because they were scattered from, for example, orbits nearer the Galactic plane.

The two-armed phase spiral remains a complicated and poorly understood phenomenon. Understanding why it is restricted to a relatively small range in angular momentum, and why it varies in shape substantially across the phase angles we are able to study, may offer us an insight into the structure and history of this region of the Galaxy where dynamical perturbations can be expected to be dominated by the Milky Way's spiral arms.

\begin{acknowledgements}
    \newline PM gratefully acknowledges support from a project grant from the Swedish Research Council (Vetenskapr\aa det, Reg: 2021-04153). TB and SA acknowledge support from project grant No.~2018-04857 from the Swedish Research Council. Some of the computations in this project were completed on computing equipment bought with a grant from The Royal Physiographic Society in Lund.
    This work has made use of data from the European Space Agency (ESA) mission
    {\it Gaia} (\url{https://www.cosmos.esa.int/gaia}), processed by the {\it Gaia}
    Data Processing and Analysis Consortium (DPAC,
    \url{https://www.cosmos.esa.int/web/gaia/dpac/consortium}). Funding for the DPAC
    has been provided by national institutions, in particular the institutions
    participating in the {\it Gaia} Multilateral Agreement.
    This research has made use of NASA’s Astrophysics Data System.
    This work made use of the following software packages for Python,
    \verb|AstroPy| \citep{astropy_collaboration_astropy_2022}, 
    \verb|emcee| \cite{foreman-mackey_emcee_2013}, 
    \verb|Numpy| \citep{harris_array_2020}, 
    \verb|Matplotlib| \citep{hunter_matplotlib_2007},
    \verb|SciPy| \citep{virtanen_scipy_2020}.
\end{acknowledgements}


\bibliographystyle{aa}
\bibliography{aanda.bib}

\begin{thebibliography}{50}
\expandafter\ifx\csname natexlab\endcsname\relax\def\natexlab#1{#1}\fi

\bibitem[{{Alinder} {et~al.}(2023){Alinder}, {McMillan}, \&
  {Bensby}}]{alinder_investigating_2023}
{Alinder}, S., {McMillan}, P.~J., \& {Bensby}, T. 2023, \aap, 678, A46

\bibitem[{{Andrae} {et~al.}(2023){Andrae}, {Rix}, \&
  {Chandra}}]{andrae_robust_2023}
{Andrae}, R., {Rix}, H.-W., \& {Chandra}, V. 2023, \apjs, 267, 8

\bibitem[{{Antoja} {et~al.}(2018){Antoja}, {Helmi}, {Romero-G{\'o}mez}, {Katz},
  {Babusiaux}, {Drimmel}, {Evans}, {Figueras}, {Poggio}, {Reyl{\'e}}, {Robin},
  {Seabroke}, \& {Soubiran}}]{antoja_dynamically_2018}
{Antoja}, T., {Helmi}, A., {Romero-G{\'o}mez}, M., {et~al.} 2018, \nat, 561,
  360

\bibitem[{{Antoja} {et~al.}(2023){Antoja}, {Ramos}, {Garc{\'\i}a-Conde},
  {Bernet}, {Laporte}, \& {Katz}}]{antoja_phase_2023}
{Antoja}, T., {Ramos}, P., {Garc{\'\i}a-Conde}, B., {et~al.} 2023, \aap, 673,
  A115

\bibitem[{{Astropy Collaboration} {et~al.}(2022){Astropy Collaboration},
  {Price-Whelan}, {Lim}, {Earl}, {Starkman}, {Bradley}, {Shupe}, {Patil},
  {Corrales}, {Brasseur}, {N{\"o}the}, {Donath}, {Tollerud}, {Morris},
  {Ginsburg}, {Vaher}, {Weaver}, {Tocknell}, {Jamieson}, {van Kerkwijk},
  {Robitaille}, {Merry}, {Bachetti}, {G{\"u}nther}, {Aldcroft},
  {Alvarado-Montes}, {Archibald}, {B{\'o}di}, {Bapat}, {Barentsen},
  {Baz{\'a}n}, {Biswas}, {Boquien}, {Burke}, {Cara}, {Cara}, {Conroy},
  {Conseil}, {Craig}, {Cross}, {Cruz}, {D'Eugenio}, {Dencheva}, {Devillepoix},
  {Dietrich}, {Eigenbrot}, {Erben}, {Ferreira}, {Foreman-Mackey}, {Fox},
  {Freij}, {Garg}, {Geda}, {Glattly}, {Gondhalekar}, {Gordon}, {Grant},
  {Greenfield}, {Groener}, {Guest}, {Gurovich}, {Handberg}, {Hart},
  {Hatfield-Dodds}, {Homeier}, {Hosseinzadeh}, {Jenness}, {Jones}, {Joseph},
  {Kalmbach}, {Karamehmetoglu}, {Ka{\l}uszy{\'n}ski}, {Kelley}, {Kern},
  {Kerzendorf}, {Koch}, {Kulumani}, {Lee}, {Ly}, {Ma}, {MacBride}, {Maljaars},
  {Muna}, {Murphy}, {Norman}, {O'Steen}, {Oman}, {Pacifici}, {Pascual},
  {Pascual-Granado}, {Patil}, {Perren}, {Pickering}, {Rastogi}, {Roulston},
  {Ryan}, {Rykoff}, {Sabater}, {Sakurikar}, {Salgado}, {Sanghi}, {Saunders},
  {Savchenko}, {Schwardt}, {Seifert-Eckert}, {Shih}, {Jain}, {Shukla}, {Sick},
  {Simpson}, {Singanamalla}, {Singer}, {Singhal}, {Sinha}, {Sip{\H{o}}cz},
  {Spitler}, {Stansby}, {Streicher}, {{\v{S}}umak}, {Swinbank}, {Taranu},
  {Tewary}, {Tremblay}, {de Val-Borro}, {Van Kooten}, {Vasovi{\'c}}, {Verma},
  {de Miranda Cardoso}, {Williams}, {Wilson}, {Winkel}, {Wood-Vasey}, {Xue},
  {Yoachim}, {Zhang}, {Zonca}, \& {Astropy Project
  Contributors}}]{astropy_collaboration_astropy_2022}
{Astropy Collaboration}, {Price-Whelan}, A.~M., {Lim}, P.~L., {et~al.} 2022,
  \apj, 935, 167

\bibitem[{{Bailer-Jones} {et~al.}(2021){Bailer-Jones}, {Rybizki}, {Fouesneau},
  {Demleitner}, \& {Andrae}}]{bailer-jones_estimating_2021}
{Bailer-Jones}, C.~A.~L., {Rybizki}, J., {Fouesneau}, M., {Demleitner}, M., \&
  {Andrae}, R. 2021, \aj, 161, 147

\bibitem[{{Banik} {et~al.}(2022){Banik}, {Weinberg}, \& {van den
  Bosch}}]{banik_comprehensive_2022}
{Banik}, U., {Weinberg}, M.~D., \& {van den Bosch}, F.~C. 2022, \apj, 935, 135

\bibitem[{{Bennett} \& {Bovy}(2019)}]{bennett_vertical_2019}
{Bennett}, M. \& {Bovy}, J. 2019, \mnras, 482, 1417

\bibitem[{{Bennett} {et~al.}(2022){Bennett}, {Bovy}, \&
  {Hunt}}]{bennett_exploring_2022}
{Bennett}, M., {Bovy}, J., \& {Hunt}, J. A.~S. 2022, \apj, 927, 131

\bibitem[{{Bensby} {et~al.}(2014){Bensby}, {Feltzing}, \&
  {Oey}}]{bensby_exploring_2014}
{Bensby}, T., {Feltzing}, S., \& {Oey}, M.~S. 2014, \aap, 562, A71

\bibitem[{{Binney} \& {Sch{\"o}nrich}(2018)}]{binney_origin_2018}
{Binney}, J. \& {Sch{\"o}nrich}, R. 2018, \mnras, 481, 1501

\bibitem[{{Binney} \& {Tremaine}(2008)}]{binney_galactic_2008}
{Binney}, J. \& {Tremaine}, S. 2008, {Galactic Dynamics: Second Edition}

\bibitem[{{Bland-Hawthorn} {et~al.}(2019){Bland-Hawthorn}, {Sharma},
  {Tepper-Garcia}, {Binney}, {Freeman}, {Hayden}, {Kos}, {De Silva}, {Ellis},
  {Lewis}, {Asplund}, {Buder}, {Casey}, {D'Orazi}, {Duong}, {Khanna}, {Lin},
  {Lind}, {Martell}, {Ness}, {Simpson}, {Zucker}, {Zwitter}, {Kafle},
  {Quillen}, {Ting}, \& {Wyse}}]{bland-hawthorn_galah_2019}
{Bland-Hawthorn}, J., {Sharma}, S., {Tepper-Garcia}, T., {et~al.} 2019, \mnras,
  486, 1167

\bibitem[{{Cao} {et~al.}(2024){Cao}, {Li}, {Sch{\"o}nrich}, \&
  {Antoja}}]{cao_radial_2024}
{Cao}, C., {Li}, Z.-Y., {Sch{\"o}nrich}, R., \& {Antoja}, T. 2024, arXiv
  e-prints, arXiv:2403.14953

\bibitem[{{Drimmel} \& {Poggio}(2018)}]{drimmel_solar_2018}
{Drimmel}, R. \& {Poggio}, E. 2018, Research Notes of the American Astronomical
  Society, 2, 210

\bibitem[{{Foreman-Mackey} {et~al.}(2013){Foreman-Mackey}, {Hogg}, {Lang}, \&
  {Goodman}}]{foreman-mackey_emcee_2013}
{Foreman-Mackey}, D., {Hogg}, D.~W., {Lang}, D., \& {Goodman}, J. 2013, \pasp,
  125, 306

\bibitem[{{Frankel} {et~al.}(2023){Frankel}, {Bovy}, {Tremaine}, \&
  {Hogg}}]{frankel_vertical_2023}
{Frankel}, N., {Bovy}, J., {Tremaine}, S., \& {Hogg}, D.~W. 2023, \mnras, 521,
  5917

\bibitem[{{Freudenreich} {et~al.}(1994){Freudenreich}, {Berriman}, {Dwek},
  {Hauser}, {Kelsall}, {Moseley}, {Silverberg}, {Sodroski}, {Toller}, \&
  {Weiland}}]{freudenreich_dirbe_1994}
{Freudenreich}, H.~T., {Berriman}, G.~B., {Dwek}, E., {et~al.} 1994, \apjl,
  429, L69

\bibitem[{{Gaia Collaboration} {et~al.}(2021{\natexlab{a}}){Gaia
  Collaboration}, {Antoja}, {McMillan}, {Kordopatis}, {Ramos}, {Helmi},
  {Balbinot}, {Cantat-Gaudin}, {Chemin}, {Figueras}, {Jordi}, {Khanna},
  {Romero-G{\'o}mez}, {Seabroke}, {Brown}, {Vallenari}, {Prusti}, {de Bruijne},
  {Babusiaux}, {Biermann}, {Creevey}, {Evans}, {Eyer}, {Hutton}, {Jansen},
  {Klioner}, {Lammers}, {Lindegren}, {Luri}, {Mignard}, {Panem}, {Pourbaix},
  {Randich}, {Sartoretti}, {Soubiran}, {Walton}, {Arenou}, {Bailer-Jones},
  {Bastian}, {Cropper}, {Drimmel}, {Katz}, {Lattanzi}, {van Leeuwen}, {Bakker},
  {Casta{\~n}eda}, {De Angeli}, {Ducourant}, {Fabricius}, {Fouesneau},
  {Fr{\'e}mat}, {Guerra}, {Guerrier}, {Guiraud}, {Jean-Antoine Piccolo},
  {Masana}, {Messineo}, {Mowlavi}, {Nicolas}, {Nienartowicz}, {Pailler},
  {Panuzzo}, {Riclet}, {Roux}, {Sordo}, {Tanga}, {Th{\'e}venin},
  {Gracia-Abril}, {Portell}, {Teyssier}, {Altmann}, {Andrae}, {Bellas-Velidis},
  {Benson}, {Berthier}, {Blomme}, {Brugaletta}, {Burgess}, {Busso}, {Carry},
  {Cellino}, {Cheek}, {Clementini}, {Damerdji}, {Davidson}, {Delchambre},
  {Dell'Oro}, {Fern{\'a}ndez-Hern{\'a}ndez}, {Galluccio}, {Garc{\'\i}a-Lario},
  {Garcia-Reinaldos}, {Gonz{\'a}lez-N{\'u}{\~n}ez}, {Gosset}, {Haigron},
  {Halbwachs}, {Hambly}, {Harrison}, {Hatzidimitriou}, {Heiter},
  {Hern{\'a}ndez}, {Hestroffer}, {Hodgkin}, {Holl}, {Jan{\ss}en}, {Jevardat de
  Fombelle}, {Jordan}, {Krone-Martins}, {Lanzafame}, {L{\"o}ffler}, {Lorca},
  {Manteiga}, {Marchal}, {Marrese}, {Moitinho}, {Mora}, {Muinonen}, {Osborne},
  {Pancino}, {Pauwels}, {Recio-Blanco}, {Richards}, {Riello}, {Rimoldini},
  {Robin}, {Roegiers}, {Rybizki}, {Sarro}, {Siopis}, {Smith}, {Sozzetti},
  {Ulla}, {Utrilla}, {van Leeuwen}, {van Reeven}, {Abbas}, {Abreu Aramburu},
  {Accart}, {Aerts}, {Aguado}, {Ajaj}, {Altavilla}, {{\'A}lvarez}, {{\'A}lvarez
  Cid-Fuentes}, {Alves}, {Anderson}, {Varela}, {Audard}, {Baines}, {Baker},
  {Balaguer-N{\'u}{\~n}ez}, {Balog}, {Barache}, {Barbato}, {Barros}, {Barstow},
  {Bartolom{\'e}}, {Bassilana}, {Bauchet}, {Baudesson-Stella}, {Becciani},
  {Bellazzini}, {Bernet}, {Bertone}, {Bianchi}, {Blanco-Cuaresma}, {Boch},
  {Bombrun}, {Bossini}, {Bouquillon}, {Bragaglia}, {Bramante}, {Breedt},
  {Bressan}, {Brouillet}, {Bucciarelli}, {Burlacu}, {Busonero}, {Butkevich},
  {Buzzi}, {Caffau}, {Cancelliere}, {C{\'a}novas}, {Carballo}, {Carlucci},
  {Carnerero}, {Carrasco}, {Casamiquela}, {Castellani}, {Castro-Ginard},
  {Castro Sampol}, {Chaoul}, {Charlot}, {Chiavassa}, {Cioni}, {Comoretto},
  {Cooper}, {Cornez}, {Cowell}, {Crifo}, {Crosta}, {Crowley}, {Dafonte},
  {Dapergolas}, {David}, {David}, {de Laverny}, {De Luise}, {De March}, {De
  Ridder}, {de Souza}, {de Teodoro}, {de Torres}, {del Peloso}, {del Pozo},
  {Delgado}, {Delgado}, {Delisle}, {Di Matteo}, {Diakite}, {Diener},
  {Distefano}, {Dolding}, {Eappachen}, {Enke}, {Esquej}, {Fabre}, {Fabrizio},
  {Faigler}, {Fedorets}, {Fernique}, {Fienga}, {Fouron}, {Fragkoudi}, {Fraile},
  {Franke}, {Gai}, {Garabato}, {Garcia-Gutierrez}, {Garc{\'\i}a-Torres},
  {Garofalo}, {Gavras}, {Gerlach}, {Geyer}, {Giacobbe}, {Gilmore}, {Girona},
  {Giuffrida}, {Gomez}, {Gonzalez-Santamaria}, {Gonz{\'a}lez-Vidal}, {Granvik},
  {Guti{\'e}rrez-S{\'a}nchez}, {Guy}, {Hauser}, {Haywood}, {Hidalgo}, {Hilger},
  {H{\l}adczuk}, {Hobbs}, {Holland}, {Huckle}, {Jasniewicz}, {Jonker},
  {Juaristi Campillo}, {Julbe}, {Karbevska}, {Kervella}, {Kochoska},
  {Kontizas}, {Korn}, {Kostrzewa-Rutkowska}, {Kruszy{\'n}ska}, {Lambert},
  {Lanza}, {Lasne}, {Le Campion}, {Le Fustec}, {Lebreton}, {Lebzelter},
  {Leccia}, {Leclerc}, {Lecoeur-Taibi}, {Liao}, {Licata}, {Lindstr{\o}m},
  {Lister}, {Livanou}, {Lobel}, {Madrero Pardo}, {Managau}, {Mann}, {Marchant},
  {Marconi}, {Marcos Santos}, {Marinoni}, {Marocco}, {Marshall}, {Martin Polo},
  {Mart{\'\i}n-Fleitas}, {Masip}, {Massari}, {Mastrobuono-Battisti}, {Mazeh},
  {Messina}, {Michalik}, {Millar}, {Mints}, {Molina}, {Molinaro}, {Moln{\'a}r},
  {Montegriffo}, {Mor}, {Morbidelli}, {Morel}, {Morris}, {Mulone}, {Munoz},
  {Muraveva}, {Murphy}, {Musella}, {Noval}, {Ord{\'e}novic}, {Orr{\`u}},
  {Osinde}, {Pagani}, {Pagano}, {Palaversa}, {Palicio}, {Panahi}, {Pawlak},
  {Pe{\~n}alosa Esteller}, {Penttil{\"a}}, {Piersimoni}, {Pineau}, {Plachy},
  {Plum}, {Poggio}, {Poretti}, {Poujoulet}, {Pr{\v{s}}a}, {Pulone}, {Racero},
  {Ragaini}, {Rainer}, {Raiteri}, {Rambaux}, {Ramos-Lerate}, {Re Fiorentin},
  {Regibo}, {Reyl{\'e}}, {Ripepi}, {Riva}, {Rixon}, {Robichon}, {Robin},
  {Roelens}, {Rohrbasser}, {Rowell}, {Royer}, {Rybicki}, {Sadowski},
  {Sagrist{\`a} Sell{\'e}s}, {Sahlmann}, {Salgado}, {Salguero}, {Samaras},
  {Sanchez Gimenez}, {Sanna}, {Santove{\~n}a}, {Sarasso}, {Schultheis},
  {Sciacca}, {Segol}, {Segovia}, {S{\'e}gransan}, {Semeux}, {Siddiqui},
  {Siebert}, {Siltala}, {Slezak}, {Smart}, {Solano}, {Solitro}, {Souami},
  {Souchay}, {Spagna}, {Spoto}, {Steele}, {Steidelm{\"u}ller}, {Stephenson},
  {S{\"u}veges}, {Szabados}, {Szegedi-Elek}, {Taris}, {Tauran}, {Taylor},
  {Teixeira}, {Thuillot}, {Tonello}, {Torra}, {Torra}, {Turon}, {Unger},
  {Vaillant}, {van Dillen}, {Vanel}, {Vecchiato}, {Viala}, {Vicente},
  {Voutsinas}, {Weiler}, {Wevers}, {Wyrzykowski}, {Yoldas}, {Yvard}, {Zhao},
  {Zorec}, {Zucker}, {Zurbach}, \& {Zwitter}}]{gaia_collaboration_gaia_2021}
{Gaia Collaboration}, {Antoja}, T., {McMillan}, P.~J., {et~al.}
  2021{\natexlab{a}}, \aap, 649, A8

\bibitem[{{Gaia Collaboration} {et~al.}(2018){Gaia Collaboration}, {Brown},
  {Vallenari}, {Prusti}, {de Bruijne}, {Babusiaux}, {Bailer-Jones}, {Biermann},
  {Evans}, {Eyer}, {Jansen}, {Jordi}, {Klioner}, {Lammers}, {Lindegren},
  {Luri}, {Mignard}, {Panem}, {Pourbaix}, {Randich}, {Sartoretti}, {Siddiqui},
  {Soubiran}, {van Leeuwen}, {Walton}, {Arenou}, {Bastian}, {Cropper},
  {Drimmel}, {Katz}, {Lattanzi}, {Bakker}, {Cacciari}, {Casta{\~n}eda},
  {Chaoul}, {Cheek}, {De Angeli}, {Fabricius}, {Guerra}, {Holl}, {Masana},
  {Messineo}, {Mowlavi}, {Nienartowicz}, {Panuzzo}, {Portell}, {Riello},
  {Seabroke}, {Tanga}, {Th{\'e}venin}, {Gracia-Abril}, {Comoretto},
  {Garcia-Reinaldos}, {Teyssier}, {Altmann}, {Andrae}, {Audard},
  {Bellas-Velidis}, {Benson}, {Berthier}, {Blomme}, {Burgess}, {Busso},
  {Carry}, {Cellino}, {Clementini}, {Clotet}, {Creevey}, {Davidson}, {De
  Ridder}, {Delchambre}, {Dell'Oro}, {Ducourant},
  {Fern{\'a}ndez-Hern{\'a}ndez}, {Fouesneau}, {Fr{\'e}mat}, {Galluccio},
  {Garc{\'\i}a-Torres}, {Gonz{\'a}lez-N{\'u}{\~n}ez}, {Gonz{\'a}lez-Vidal},
  {Gosset}, {Guy}, {Halbwachs}, {Hambly}, {Harrison}, {Hern{\'a}ndez},
  {Hestroffer}, {Hodgkin}, {Hutton}, {Jasniewicz}, {Jean-Antoine-Piccolo},
  {Jordan}, {Korn}, {Krone-Martins}, {Lanzafame}, {Lebzelter}, {L{\"o}ffler},
  {Manteiga}, {Marrese}, {Mart{\'\i}n-Fleitas}, {Moitinho}, {Mora}, {Muinonen},
  {Osinde}, {Pancino}, {Pauwels}, {Petit}, {Recio-Blanco}, {Richards},
  {Rimoldini}, {Robin}, {Sarro}, {Siopis}, {Smith}, {Sozzetti}, {S{\"u}veges},
  {Torra}, {van Reeven}, {Abbas}, {Abreu Aramburu}, {Accart}, {Aerts},
  {Altavilla}, {{\'A}lvarez}, {Alvarez}, {Alves}, {Anderson}, {Andrei},
  {Anglada Varela}, {Antiche}, {Antoja}, {Arcay}, {Astraatmadja}, {Bach},
  {Baker}, {Balaguer-N{\'u}{\~n}ez}, {Balm}, {Barache}, {Barata}, {Barbato},
  {Barblan}, {Barklem}, {Barrado}, {Barros}, {Barstow}, {Bartholom{\'e}
  Mu{\~n}oz}, {Bassilana}, {Becciani}, {Bellazzini}, {Berihuete}, {Bertone},
  {Bianchi}, {Bienaym{\'e}}, {Blanco-Cuaresma}, {Boch}, {Boeche}, {Bombrun},
  {Borrachero}, {Bossini}, {Bouquillon}, {Bourda}, {Bragaglia}, {Bramante},
  {Breddels}, {Bressan}, {Brouillet}, {Br{\"u}semeister}, {Brugaletta},
  {Bucciarelli}, {Burlacu}, {Busonero}, {Butkevich}, {Buzzi}, {Caffau},
  {Cancelliere}, {Cannizzaro}, {Cantat-Gaudin}, {Carballo}, {Carlucci},
  {Carrasco}, {Casamiquela}, {Castellani}, {Castro-Ginard}, {Charlot},
  {Chemin}, {Chiavassa}, {Cocozza}, {Costigan}, {Cowell}, {Crifo}, {Crosta},
  {Crowley}, {Cuypers}, {Dafonte}, {Damerdji}, {Dapergolas}, {David}, {David},
  {de Laverny}, {De Luise}, {De March}, {de Martino}, {de Souza}, {de Torres},
  {Debosscher}, {del Pozo}, {Delbo}, {Delgado}, {Delgado}, {Di Matteo},
  {Diakite}, {Diener}, {Distefano}, {Dolding}, {Drazinos}, {Dur{\'a}n},
  {Edvardsson}, {Enke}, {Eriksson}, {Esquej}, {Eynard Bontemps}, {Fabre},
  {Fabrizio}, {Faigler}, {Falc{\~a}o}, {Farr{\`a}s Casas}, {Federici},
  {Fedorets}, {Fernique}, {Figueras}, {Filippi}, {Findeisen}, {Fonti},
  {Fraile}, {Fraser}, {Fr{\'e}zouls}, {Gai}, {Galleti}, {Garabato},
  {Garc{\'\i}a-Sedano}, {Garofalo}, {Garralda}, {Gavel}, {Gavras}, {Gerssen},
  {Geyer}, {Giacobbe}, {Gilmore}, {Girona}, {Giuffrida}, {Glass}, {Gomes},
  {Granvik}, {Gueguen}, {Guerrier}, {Guiraud}, {Guti{\'e}rrez-S{\'a}nchez},
  {Haigron}, {Hatzidimitriou}, {Hauser}, {Haywood}, {Heiter}, {Helmi}, {Heu},
  {Hilger}, {Hobbs}, {Hofmann}, {Holland}, {Huckle}, {Hypki}, {Icardi},
  {Jan{\ss}en}, {Jevardat de Fombelle}, {Jonker}, {Juh{\'a}sz}, {Julbe},
  {Karampelas}, {Kewley}, {Klar}, {Kochoska}, {Kohley}, {Kolenberg},
  {Kontizas}, {Kontizas}, {Koposov}, {Kordopatis}, {Kostrzewa-Rutkowska},
  {Koubsky}, {Lambert}, {Lanza}, {Lasne}, {Lavigne}, {Le Fustec}, {Le
  Poncin-Lafitte}, {Lebreton}, {Leccia}, {Leclerc}, {Lecoeur-Taibi},
  {Lenhardt}, {Leroux}, {Liao}, {Licata}, {Lindstr{\o}m}, {Lister}, {Livanou},
  {Lobel}, {L{\'o}pez}, {Managau}, {Mann}, {Mantelet}, {Marchal}, {Marchant},
  {Marconi}, {Marinoni}, {Marschalk{\'o}}, {Marshall}, {Martino}, {Marton},
  {Mary}, {Massari}, {Matijevi{\v{c}}}, {Mazeh}, {McMillan}, {Messina},
  {Michalik}, {Millar}, {Molina}, {Molinaro}, {Moln{\'a}r}, {Montegriffo},
  {Mor}, {Morbidelli}, {Morel}, {Morris}, {Mulone}, {Muraveva}, {Musella},
  {Nelemans}, {Nicastro}, {Noval}, {O'Mullane}, {Ord{\'e}novic},
  {Ord{\'o}{\~n}ez-Blanco}, {Osborne}, {Pagani}, {Pagano}, {Pailler},
  {Palacin}, {Palaversa}, {Panahi}, {Pawlak}, {Piersimoni}, {Pineau}, {Plachy},
  {Plum}, {Poggio}, {Poujoulet}, {Pr{\v{s}}a}, {Pulone}, {Racero}, {Ragaini},
  {Rambaux}, {Ramos-Lerate}, {Regibo}, {Reyl{\'e}}, {Riclet}, {Ripepi}, {Riva},
  {Rivard}, {Rixon}, {Roegiers}, {Roelens}, {Romero-G{\'o}mez}, {Rowell},
  {Royer}, {Ruiz-Dern}, {Sadowski}, {Sagrist{\`a} Sell{\'e}s}, {Sahlmann},
  {Salgado}, {Salguero}, {Sanna}, {Santana-Ros}, {Sarasso}, {Savietto},
  {Schultheis}, {Sciacca}, {Segol}, {Segovia}, {S{\'e}gransan}, {Shih},
  {Siltala}, {Silva}, {Smart}, {Smith}, {Solano}, {Solitro}, {Sordo}, {Soria
  Nieto}, {Souchay}, {Spagna}, {Spoto}, {Stampa}, {Steele},
  {Steidelm{\"u}ller}, {Stephenson}, {Stoev}, {Suess}, {Surdej}, {Szabados},
  {Szegedi-Elek}, {Tapiador}, {Taris}, {Tauran}, {Taylor}, {Teixeira},
  {Terrett}, {Teyssandier}, {Thuillot}, {Titarenko}, {Torra Clotet}, {Turon},
  {Ulla}, {Utrilla}, {Uzzi}, {Vaillant}, {Valentini}, {Valette}, {van Elteren},
  {Van Hemelryck}, {van Leeuwen}, {Vaschetto}, {Vecchiato}, {Veljanoski},
  {Viala}, {Vicente}, {Vogt}, {von Essen}, {Voss}, {Votruba}, {Voutsinas},
  {Walmsley}, {Weiler}, {Wertz}, {Wevers}, {Wyrzykowski}, {Yoldas},
  {{\v{Z}}erjal}, {Ziaeepour}, {Zorec}, {Zschocke}, {Zucker}, {Zurbach}, \&
  {Zwitter}}]{gaia_collaboration_gaia_2018}
{Gaia Collaboration}, {Brown}, A.~G.~A., {Vallenari}, A., {et~al.} 2018, \aap,
  616, A1

\bibitem[{{Gaia Collaboration} {et~al.}(2021{\natexlab{b}}){Gaia
  Collaboration}, {Brown}, {Vallenari}, {Prusti}, {de Bruijne}, {Babusiaux},
  {Biermann}, {Creevey}, {Evans}, {Eyer}, {Hutton}, {Jansen}, {Jordi},
  {Klioner}, {Lammers}, {Lindegren}, {Luri}, {Mignard}, {Panem}, {Pourbaix},
  {Randich}, {Sartoretti}, {Soubiran}, {Walton}, {Arenou}, {Bailer-Jones},
  {Bastian}, {Cropper}, {Drimmel}, {Katz}, {Lattanzi}, {van Leeuwen}, {Bakker},
  {Cacciari}, {Casta{\~n}eda}, {De Angeli}, {Ducourant}, {Fabricius},
  {Fouesneau}, {Fr{\'e}mat}, {Guerra}, {Guerrier}, {Guiraud}, {Jean-Antoine
  Piccolo}, {Masana}, {Messineo}, {Mowlavi}, {Nicolas}, {Nienartowicz},
  {Pailler}, {Panuzzo}, {Riclet}, {Roux}, {Seabroke}, {Sordo}, {Tanga},
  {Th{\'e}venin}, {Gracia-Abril}, {Portell}, {Teyssier}, {Altmann}, {Andrae},
  {Bellas-Velidis}, {Benson}, {Berthier}, {Blomme}, {Brugaletta}, {Burgess},
  {Busso}, {Carry}, {Cellino}, {Cheek}, {Clementini}, {Damerdji}, {Davidson},
  {Delchambre}, {Dell'Oro}, {Fern{\'a}ndez-Hern{\'a}ndez}, {Galluccio},
  {Garc{\'\i}a-Lario}, {Garcia-Reinaldos}, {Gonz{\'a}lez-N{\'u}{\~n}ez},
  {Gosset}, {Haigron}, {Halbwachs}, {Hambly}, {Harrison}, {Hatzidimitriou},
  {Heiter}, {Hern{\'a}ndez}, {Hestroffer}, {Hodgkin}, {Holl}, {Jan{\ss}en},
  {Jevardat de Fombelle}, {Jordan}, {Krone-Martins}, {Lanzafame},
  {L{\"o}ffler}, {Lorca}, {Manteiga}, {Marchal}, {Marrese}, {Moitinho}, {Mora},
  {Muinonen}, {Osborne}, {Pancino}, {Pauwels}, {Petit}, {Recio-Blanco},
  {Richards}, {Riello}, {Rimoldini}, {Robin}, {Roegiers}, {Rybizki}, {Sarro},
  {Siopis}, {Smith}, {Sozzetti}, {Ulla}, {Utrilla}, {van Leeuwen}, {van
  Reeven}, {Abbas}, {Abreu Aramburu}, {Accart}, {Aerts}, {Aguado}, {Ajaj},
  {Altavilla}, {{\'A}lvarez}, {{\'A}lvarez Cid-Fuentes}, {Alves}, {Anderson},
  {Anglada Varela}, {Antoja}, {Audard}, {Baines}, {Baker},
  {Balaguer-N{\'u}{\~n}ez}, {Balbinot}, {Balog}, {Barache}, {Barbato},
  {Barros}, {Barstow}, {Bartolom{\'e}}, {Bassilana}, {Bauchet},
  {Baudesson-Stella}, {Becciani}, {Bellazzini}, {Bernet}, {Bertone}, {Bianchi},
  {Blanco-Cuaresma}, {Boch}, {Bombrun}, {Bossini}, {Bouquillon}, {Bragaglia},
  {Bramante}, {Breedt}, {Bressan}, {Brouillet}, {Bucciarelli}, {Burlacu},
  {Busonero}, {Butkevich}, {Buzzi}, {Caffau}, {Cancelliere}, {C{\'a}novas},
  {Cantat-Gaudin}, {Carballo}, {Carlucci}, {Carnerero}, {Carrasco},
  {Casamiquela}, {Castellani}, {Castro-Ginard}, {Castro Sampol}, {Chaoul},
  {Charlot}, {Chemin}, {Chiavassa}, {Cioni}, {Comoretto}, {Cooper}, {Cornez},
  {Cowell}, {Crifo}, {Crosta}, {Crowley}, {Dafonte}, {Dapergolas}, {David},
  {David}, {de Laverny}, {De Luise}, {De March}, {De Ridder}, {de Souza}, {de
  Teodoro}, {de Torres}, {del Peloso}, {del Pozo}, {Delbo}, {Delgado},
  {Delgado}, {Delisle}, {Di Matteo}, {Diakite}, {Diener}, {Distefano},
  {Dolding}, {Eappachen}, {Edvardsson}, {Enke}, {Esquej}, {Fabre}, {Fabrizio},
  {Faigler}, {Fedorets}, {Fernique}, {Fienga}, {Figueras}, {Fouron},
  {Fragkoudi}, {Fraile}, {Franke}, {Gai}, {Garabato}, {Garcia-Gutierrez},
  {Garc{\'\i}a-Torres}, {Garofalo}, {Gavras}, {Gerlach}, {Geyer}, {Giacobbe},
  {Gilmore}, {Girona}, {Giuffrida}, {Gomel}, {Gomez}, {Gonzalez-Santamaria},
  {Gonz{\'a}lez-Vidal}, {Granvik}, {Guti{\'e}rrez-S{\'a}nchez}, {Guy},
  {Hauser}, {Haywood}, {Helmi}, {Hidalgo}, {Hilger}, {H{\l}adczuk}, {Hobbs},
  {Holland}, {Huckle}, {Jasniewicz}, {Jonker}, {Juaristi Campillo}, {Julbe},
  {Karbevska}, {Kervella}, {Khanna}, {Kochoska}, {Kontizas}, {Kordopatis},
  {Korn}, {Kostrzewa-Rutkowska}, {Kruszy{\'n}ska}, {Lambert}, {Lanza}, {Lasne},
  {Le Campion}, {Le Fustec}, {Lebreton}, {Lebzelter}, {Leccia}, {Leclerc},
  {Lecoeur-Taibi}, {Liao}, {Licata}, {Lindstr{\o}m}, {Lister}, {Livanou},
  {Lobel}, {Madrero Pardo}, {Managau}, {Mann}, {Marchant}, {Marconi}, {Marcos
  Santos}, {Marinoni}, {Marocco}, {Marshall}, {Martin Polo},
  {Mart{\'\i}n-Fleitas}, {Masip}, {Massari}, {Mastrobuono-Battisti}, {Mazeh},
  {McMillan}, {Messina}, {Michalik}, {Millar}, {Mints}, {Molina}, {Molinaro},
  {Moln{\'a}r}, {Montegriffo}, {Mor}, {Morbidelli}, {Morel}, {Morris},
  {Mulone}, {Munoz}, {Muraveva}, {Murphy}, {Musella}, {Noval}, {Ord{\'e}novic},
  {Orr{\`u}}, {Osinde}, {Pagani}, {Pagano}, {Palaversa}, {Palicio}, {Panahi},
  {Pawlak}, {Pe{\~n}alosa Esteller}, {Penttil{\"a}}, {Piersimoni}, {Pineau},
  {Plachy}, {Plum}, {Poggio}, {Poretti}, {Poujoulet}, {Pr{\v{s}}a}, {Pulone},
  {Racero}, {Ragaini}, {Rainer}, {Raiteri}, {Rambaux}, {Ramos}, {Ramos-Lerate},
  {Re Fiorentin}, {Regibo}, {Reyl{\'e}}, {Ripepi}, {Riva}, {Rixon}, {Robichon},
  {Robin}, {Roelens}, {Rohrbasser}, {Romero-G{\'o}mez}, {Rowell}, {Royer},
  {Rybicki}, {Sadowski}, {Sagrist{\`a} Sell{\'e}s}, {Sahlmann}, {Salgado},
  {Salguero}, {Samaras}, {Sanchez Gimenez}, {Sanna}, {Santove{\~n}a},
  {Sarasso}, {Schultheis}, {Sciacca}, {Segol}, {Segovia}, {S{\'e}gransan},
  {Semeux}, {Shahaf}, {Siddiqui}, {Siebert}, {Siltala}, {Slezak}, {Smart},
  {Solano}, {Solitro}, {Souami}, {Souchay}, {Spagna}, {Spoto}, {Steele},
  {Steidelm{\"u}ller}, {Stephenson}, {S{\"u}veges}, {Szabados}, {Szegedi-Elek},
  {Taris}, {Tauran}, {Taylor}, {Teixeira}, {Thuillot}, {Tonello}, {Torra},
  {Torra}, {Turon}, {Unger}, {Vaillant}, {van Dillen}, {Vanel}, {Vecchiato},
  {Viala}, {Vicente}, {Voutsinas}, {Weiler}, {Wevers}, {Wyrzykowski}, {Yoldas},
  {Yvard}, {Zhao}, {Zorec}, {Zucker}, {Zurbach}, \&
  {Zwitter}}]{gaia_collaboration_gaia_2021-1}
{Gaia Collaboration}, {Brown}, A.~G.~A., {Vallenari}, A., {et~al.}
  2021{\natexlab{b}}, \aap, 649, A1

\bibitem[{{Gaia Collaboration} {et~al.}(2016{\natexlab{a}}){Gaia
  Collaboration}, {Brown}, {Vallenari}, {Prusti}, {de Bruijne}, {Mignard},
  {Drimmel}, {Babusiaux}, {Bailer-Jones}, {Bastian}, {Biermann}, {Evans},
  {Eyer}, {Jansen}, {Jordi}, {Katz}, {Klioner}, {Lammers}, {Lindegren}, {Luri},
  {O'Mullane}, {Panem}, {Pourbaix}, {Randich}, {Sartoretti}, {Siddiqui},
  {Soubiran}, {Valette}, {van Leeuwen}, {Walton}, {Aerts}, {Arenou}, {Cropper},
  {H{\o}g}, {Lattanzi}, {Grebel}, {Holland}, {Huc}, {Passot}, {Perryman},
  {Bramante}, {Cacciari}, {Casta{\~n}eda}, {Chaoul}, {Cheek}, {De Angeli},
  {Fabricius}, {Guerra}, {Hern{\'a}ndez}, {Jean-Antoine-Piccolo}, {Masana},
  {Messineo}, {Mowlavi}, {Nienartowicz}, {Ord{\'o}{\~n}ez-Blanco}, {Panuzzo},
  {Portell}, {Richards}, {Riello}, {Seabroke}, {Tanga}, {Th{\'e}venin},
  {Torra}, {Els}, {Gracia-Abril}, {Comoretto}, {Garcia-Reinaldos}, {Lock},
  {Mercier}, {Altmann}, {Andrae}, {Astraatmadja}, {Bellas-Velidis}, {Benson},
  {Berthier}, {Blomme}, {Busso}, {Carry}, {Cellino}, {Clementini}, {Cowell},
  {Creevey}, {Cuypers}, {Davidson}, {De Ridder}, {de Torres}, {Delchambre},
  {Dell'Oro}, {Ducourant}, {Fr{\'e}mat}, {Garc{\'\i}a-Torres}, {Gosset},
  {Halbwachs}, {Hambly}, {Harrison}, {Hauser}, {Hestroffer}, {Hodgkin},
  {Huckle}, {Hutton}, {Jasniewicz}, {Jordan}, {Kontizas}, {Korn}, {Lanzafame},
  {Manteiga}, {Moitinho}, {Muinonen}, {Osinde}, {Pancino}, {Pauwels}, {Petit},
  {Recio-Blanco}, {Robin}, {Sarro}, {Siopis}, {Smith}, {Smith}, {Sozzetti},
  {Thuillot}, {van Reeven}, {Viala}, {Abbas}, {Abreu Aramburu}, {Accart},
  {Aguado}, {Allan}, {Allasia}, {Altavilla}, {{\'A}lvarez}, {Alves},
  {Anderson}, {Andrei}, {Anglada Varela}, {Antiche}, {Antoja}, {Ant{\'o}n},
  {Arcay}, {Bach}, {Baker}, {Balaguer-N{\'u}{\~n}ez}, {Barache}, {Barata},
  {Barbier}, {Barblan}, {Barrado y Navascu{\'e}s}, {Barros}, {Barstow},
  {Becciani}, {Bellazzini}, {Bello Garc{\'\i}a}, {Belokurov}, {Bendjoya},
  {Berihuete}, {Bianchi}, {Bienaym{\'e}}, {Billebaud}, {Blagorodnova},
  {Blanco-Cuaresma}, {Boch}, {Bombrun}, {Borrachero}, {Bouquillon}, {Bourda},
  {Bouy}, {Bragaglia}, {Breddels}, {Brouillet}, {Br{\"u}semeister},
  {Bucciarelli}, {Burgess}, {Burgon}, {Burlacu}, {Busonero}, {Buzzi}, {Caffau},
  {Cambras}, {Campbell}, {Cancelliere}, {Cantat-Gaudin}, {Carlucci},
  {Carrasco}, {Castellani}, {Charlot}, {Charnas}, {Chiavassa}, {Clotet},
  {Cocozza}, {Collins}, {Costigan}, {Crifo}, {Cross}, {Crosta}, {Crowley},
  {Dafonte}, {Damerdji}, {Dapergolas}, {David}, {David}, {De Cat}, {de Felice},
  {de Laverny}, {De Luise}, {De March}, {de Martino}, {de Souza}, {Debosscher},
  {del Pozo}, {Delbo}, {Delgado}, {Delgado}, {Di Matteo}, {Diakite},
  {Distefano}, {Dolding}, {Dos Anjos}, {Drazinos}, {Duran}, {Dzigan},
  {Edvardsson}, {Enke}, {Evans}, {Eynard Bontemps}, {Fabre}, {Fabrizio},
  {Faigler}, {Falc{\~a}o}, {Farr{\`a}s Casas}, {Federici}, {Fedorets},
  {Fern{\'a}ndez-Hern{\'a}ndez}, {Fernique}, {Fienga}, {Figueras}, {Filippi},
  {Findeisen}, {Fonti}, {Fouesneau}, {Fraile}, {Fraser}, {Fuchs}, {Gai},
  {Galleti}, {Galluccio}, {Garabato}, {Garc{\'\i}a-Sedano}, {Garofalo},
  {Garralda}, {Gavras}, {Gerssen}, {Geyer}, {Gilmore}, {Girona}, {Giuffrida},
  {Gomes}, {Gonz{\'a}lez-Marcos}, {Gonz{\'a}lez-N{\'u}{\~n}ez},
  {Gonz{\'a}lez-Vidal}, {Granvik}, {Guerrier}, {Guillout}, {Guiraud},
  {G{\'u}rpide}, {Guti{\'e}rrez-S{\'a}nchez}, {Guy}, {Haigron},
  {Hatzidimitriou}, {Haywood}, {Heiter}, {Helmi}, {Hobbs}, {Hofmann}, {Holl},
  {Holland}, {Hunt}, {Hypki}, {Icardi}, {Irwin}, {Jevardat de Fombelle},
  {Jofr{\'e}}, {Jonker}, {Jorissen}, {Julbe}, {Karampelas}, {Kochoska},
  {Kohley}, {Kolenberg}, {Kontizas}, {Koposov}, {Kordopatis}, {Koubsky},
  {Krone-Martins}, {Kudryashova}, {Kull}, {Bachchan}, {Lacoste-Seris}, {Lanza},
  {Lavigne}, {Le Poncin-Lafitte}, {Lebreton}, {Lebzelter}, {Leccia}, {Leclerc},
  {Lecoeur-Taibi}, {Lemaitre}, {Lenhardt}, {Leroux}, {Liao}, {Licata},
  {Lindstr{\o}m}, {Lister}, {Livanou}, {Lobel}, {L{\"o}ffler}, {L{\'o}pez},
  {Lorenz}, {MacDonald}, {Magalh{\~a}es Fernandes}, {Managau}, {Mann},
  {Mantelet}, {Marchal}, {Marchant}, {Marconi}, {Marinoni}, {Marrese},
  {Marschalk{\'o}}, {Marshall}, {Mart{\'\i}n-Fleitas}, {Martino}, {Mary},
  {Matijevi{\v{c}}}, {Mazeh}, {McMillan}, {Messina}, {Michalik}, {Millar},
  {Miranda}, {Molina}, {Molinaro}, {Molinaro}, {Moln{\'a}r}, {Moniez},
  {Montegriffo}, {Mor}, {Mora}, {Morbidelli}, {Morel}, {Morgenthaler},
  {Morris}, {Mulone}, {Muraveva}, {Musella}, {Narbonne}, {Nelemans},
  {Nicastro}, {Noval}, {Ord{\'e}novic}, {Ordieres-Mer{\'e}}, {Osborne},
  {Pagani}, {Pagano}, {Pailler}, {Palacin}, {Palaversa}, {Parsons}, {Pecoraro},
  {Pedrosa}, {Pentik{\"a}inen}, {Pichon}, {Piersimoni}, {Pineau}, {Plachy},
  {Plum}, {Poujoulet}, {Pr{\v{s}}a}, {Pulone}, {Ragaini}, {Rago}, {Rambaux},
  {Ramos-Lerate}, {Ranalli}, {Rauw}, {Read}, {Regibo}, {Reyl{\'e}}, {Ribeiro},
  {Rimoldini}, {Ripepi}, {Riva}, {Rixon}, {Roelens}, {Romero-G{\'o}mez},
  {Rowell}, {Royer}, {Ruiz-Dern}, {Sadowski}, {Sagrist{\`a} Sell{\'e}s},
  {Sahlmann}, {Salgado}, {Salguero}, {Sarasso}, {Savietto}, {Schultheis},
  {Sciacca}, {Segol}, {Segovia}, {Segransan}, {Shih}, {Smareglia}, {Smart},
  {Solano}, {Solitro}, {Sordo}, {Soria Nieto}, {Souchay}, {Spagna}, {Spoto},
  {Stampa}, {Steele}, {Steidelm{\"u}ller}, {Stephenson}, {Stoev}, {Suess},
  {S{\"u}veges}, {Surdej}, {Szabados}, {Szegedi-Elek}, {Tapiador}, {Taris},
  {Tauran}, {Taylor}, {Teixeira}, {Terrett}, {Tingley}, {Trager}, {Turon},
  {Ulla}, {Utrilla}, {Valentini}, {van Elteren}, {Van Hemelryck}, {van
  Leeuwen}, {Varadi}, {Vecchiato}, {Veljanoski}, {Via}, {Vicente}, {Vogt},
  {Voss}, {Votruba}, {Voutsinas}, {Walmsley}, {Weiler}, {Weingrill}, {Wevers},
  {Wyrzykowski}, {Yoldas}, {{\v{Z}}erjal}, {Zucker}, {Zurbach}, {Zwitter},
  {Alecu}, {Allen}, {Allende Prieto}, {Amorim}, {Anglada-Escud{\'e}},
  {Arsenijevic}, {Azaz}, {Balm}, {Beck}, {Bernstein}, {Bigot}, {Bijaoui},
  {Blasco}, {Bonfigli}, {Bono}, {Boudreault}, {Bressan}, {Brown}, {Brunet},
  {Bunclark}, {Buonanno}, {Butkevich}, {Carret}, {Carrion}, {Chemin},
  {Ch{\'e}reau}, {Corcione}, {Darmigny}, {de Boer}, {de Teodoro}, {de Zeeuw},
  {Delle Luche}, {Domingues}, {Dubath}, {Fodor}, {Fr{\'e}zouls}, {Fries},
  {Fustes}, {Fyfe}, {Gallardo}, {Gallegos}, {Gardiol}, {Gebran}, {Gomboc},
  {G{\'o}mez}, {Grux}, {Gueguen}, {Heyrovsky}, {Hoar}, {Iannicola}, {Isasi
  Parache}, {Janotto}, {Joliet}, {Jonckheere}, {Keil}, {Kim}, {Klagyivik},
  {Klar}, {Knude}, {Kochukhov}, {Kolka}, {Kos}, {Kutka}, {Lainey}, {LeBouquin},
  {Liu}, {Loreggia}, {Makarov}, {Marseille}, {Martayan}, {Martinez-Rubi},
  {Massart}, {Meynadier}, {Mignot}, {Munari}, {Nguyen}, {Nordlander}, {Ocvirk},
  {O'Flaherty}, {Olias Sanz}, {Ortiz}, {Osorio}, {Oszkiewicz}, {Ouzounis},
  {Palmer}, {Park}, {Pasquato}, {Peltzer}, {Peralta}, {P{\'e}turaud},
  {Pieniluoma}, {Pigozzi}, {Poels}, {Prat}, {Prod'homme}, {Raison}, {Rebordao},
  {Risquez}, {Rocca-Volmerange}, {Rosen}, {Ruiz-Fuertes}, {Russo}, {Sembay},
  {Serraller Vizcaino}, {Short}, {Siebert}, {Silva}, {Sinachopoulos}, {Slezak},
  {Soffel}, {Sosnowska}, {Strai{\v{z}}ys}, {ter Linden}, {Terrell}, {Theil},
  {Tiede}, {Troisi}, {Tsalmantza}, {Tur}, {Vaccari}, {Vachier}, {Valles}, {Van
  Hamme}, {Veltz}, {Virtanen}, {Wallut}, {Wichmann}, {Wilkinson}, {Ziaeepour},
  \& {Zschocke}}]{gaia_collaboration_gaia_2016}
{Gaia Collaboration}, {Brown}, A.~G.~A., {Vallenari}, A., {et~al.}
  2016{\natexlab{a}}, \aap, 595, A2

\bibitem[{{Gaia Collaboration} {et~al.}(2016{\natexlab{b}}){Gaia
  Collaboration}, {Prusti}, {de Bruijne}, {Brown}, {Vallenari}, {Babusiaux},
  {Bailer-Jones}, {Bastian}, {Biermann}, {Evans}, {Eyer}, {Jansen}, {Jordi},
  {Klioner}, {Lammers}, {Lindegren}, {Luri}, {Mignard}, {Milligan}, {Panem},
  {Poinsignon}, {Pourbaix}, {Randich}, {Sarri}, {Sartoretti}, {Siddiqui},
  {Soubiran}, {Valette}, {van Leeuwen}, {Walton}, {Aerts}, {Arenou}, {Cropper},
  {Drimmel}, {H{\o}g}, {Katz}, {Lattanzi}, {O'Mullane}, {Grebel}, {Holland},
  {Huc}, {Passot}, {Bramante}, {Cacciari}, {Casta{\~n}eda}, {Chaoul}, {Cheek},
  {De Angeli}, {Fabricius}, {Guerra}, {Hern{\'a}ndez}, {Jean-Antoine-Piccolo},
  {Masana}, {Messineo}, {Mowlavi}, {Nienartowicz}, {Ord{\'o}{\~n}ez-Blanco},
  {Panuzzo}, {Portell}, {Richards}, {Riello}, {Seabroke}, {Tanga},
  {Th{\'e}venin}, {Torra}, {Els}, {Gracia-Abril}, {Comoretto},
  {Garcia-Reinaldos}, {Lock}, {Mercier}, {Altmann}, {Andrae}, {Astraatmadja},
  {Bellas-Velidis}, {Benson}, {Berthier}, {Blomme}, {Busso}, {Carry},
  {Cellino}, {Clementini}, {Cowell}, {Creevey}, {Cuypers}, {Davidson}, {De
  Ridder}, {de Torres}, {Delchambre}, {Dell'Oro}, {Ducourant}, {Fr{\'e}mat},
  {Garc{\'\i}a-Torres}, {Gosset}, {Halbwachs}, {Hambly}, {Harrison}, {Hauser},
  {Hestroffer}, {Hodgkin}, {Huckle}, {Hutton}, {Jasniewicz}, {Jordan},
  {Kontizas}, {Korn}, {Lanzafame}, {Manteiga}, {Moitinho}, {Muinonen},
  {Osinde}, {Pancino}, {Pauwels}, {Petit}, {Recio-Blanco}, {Robin}, {Sarro},
  {Siopis}, {Smith}, {Smith}, {Sozzetti}, {Thuillot}, {van Reeven}, {Viala},
  {Abbas}, {Abreu Aramburu}, {Accart}, {Aguado}, {Allan}, {Allasia},
  {Altavilla}, {{\'A}lvarez}, {Alves}, {Anderson}, {Andrei}, {Anglada Varela},
  {Antiche}, {Antoja}, {Ant{\'o}n}, {Arcay}, {Atzei}, {Ayache}, {Bach},
  {Baker}, {Balaguer-N{\'u}{\~n}ez}, {Barache}, {Barata}, {Barbier}, {Barblan},
  {Baroni}, {Barrado y Navascu{\'e}s}, {Barros}, {Barstow}, {Becciani},
  {Bellazzini}, {Bellei}, {Bello Garc{\'\i}a}, {Belokurov}, {Bendjoya},
  {Berihuete}, {Bianchi}, {Bienaym{\'e}}, {Billebaud}, {Blagorodnova},
  {Blanco-Cuaresma}, {Boch}, {Bombrun}, {Borrachero}, {Bouquillon}, {Bourda},
  {Bouy}, {Bragaglia}, {Breddels}, {Brouillet}, {Br{\"u}semeister},
  {Bucciarelli}, {Budnik}, {Burgess}, {Burgon}, {Burlacu}, {Busonero}, {Buzzi},
  {Caffau}, {Cambras}, {Campbell}, {Cancelliere}, {Cantat-Gaudin}, {Carlucci},
  {Carrasco}, {Castellani}, {Charlot}, {Charnas}, {Charvet}, {Chassat},
  {Chiavassa}, {Clotet}, {Cocozza}, {Collins}, {Collins}, {Costigan}, {Crifo},
  {Cross}, {Crosta}, {Crowley}, {Dafonte}, {Damerdji}, {Dapergolas}, {David},
  {David}, {De Cat}, {de Felice}, {de Laverny}, {De Luise}, {De March}, {de
  Martino}, {de Souza}, {Debosscher}, {del Pozo}, {Delbo}, {Delgado},
  {Delgado}, {di Marco}, {Di Matteo}, {Diakite}, {Distefano}, {Dolding}, {Dos
  Anjos}, {Drazinos}, {Dur{\'a}n}, {Dzigan}, {Ecale}, {Edvardsson}, {Enke},
  {Erdmann}, {Escolar}, {Espina}, {Evans}, {Eynard Bontemps}, {Fabre},
  {Fabrizio}, {Faigler}, {Falc{\~a}o}, {Farr{\`a}s Casas}, {Faye}, {Federici},
  {Fedorets}, {Fern{\'a}ndez-Hern{\'a}ndez}, {Fernique}, {Fienga}, {Figueras},
  {Filippi}, {Findeisen}, {Fonti}, {Fouesneau}, {Fraile}, {Fraser}, {Fuchs},
  {Furnell}, {Gai}, {Galleti}, {Galluccio}, {Garabato}, {Garc{\'\i}a-Sedano},
  {Gar{\'e}}, {Garofalo}, {Garralda}, {Gavras}, {Gerssen}, {Geyer}, {Gilmore},
  {Girona}, {Giuffrida}, {Gomes}, {Gonz{\'a}lez-Marcos},
  {Gonz{\'a}lez-N{\'u}{\~n}ez}, {Gonz{\'a}lez-Vidal}, {Granvik}, {Guerrier},
  {Guillout}, {Guiraud}, {G{\'u}rpide}, {Guti{\'e}rrez-S{\'a}nchez}, {Guy},
  {Haigron}, {Hatzidimitriou}, {Haywood}, {Heiter}, {Helmi}, {Hobbs},
  {Hofmann}, {Holl}, {Holland}, {Hunt}, {Hypki}, {Icardi}, {Irwin}, {Jevardat
  de Fombelle}, {Jofr{\'e}}, {Jonker}, {Jorissen}, {Julbe}, {Karampelas},
  {Kochoska}, {Kohley}, {Kolenberg}, {Kontizas}, {Koposov}, {Kordopatis},
  {Koubsky}, {Kowalczyk}, {Krone-Martins}, {Kudryashova}, {Kull}, {Bachchan},
  {Lacoste-Seris}, {Lanza}, {Lavigne}, {Le Poncin-Lafitte}, {Lebreton},
  {Lebzelter}, {Leccia}, {Leclerc}, {Lecoeur-Taibi}, {Lemaitre}, {Lenhardt},
  {Leroux}, {Liao}, {Licata}, {Lindstr{\o}m}, {Lister}, {Livanou}, {Lobel},
  {L{\"o}ffler}, {L{\'o}pez}, {Lopez-Lozano}, {Lorenz}, {Loureiro},
  {MacDonald}, {Magalh{\~a}es Fernandes}, {Managau}, {Mann}, {Mantelet},
  {Marchal}, {Marchant}, {Marconi}, {Marie}, {Marinoni}, {Marrese},
  {Marschalk{\'o}}, {Marshall}, {Mart{\'\i}n-Fleitas}, {Martino}, {Mary},
  {Matijevi{\v{c}}}, {Mazeh}, {McMillan}, {Messina}, {Mestre}, {Michalik},
  {Millar}, {Miranda}, {Molina}, {Molinaro}, {Molinaro}, {Moln{\'a}r},
  {Moniez}, {Montegriffo}, {Monteiro}, {Mor}, {Mora}, {Morbidelli}, {Morel},
  {Morgenthaler}, {Morley}, {Morris}, {Mulone}, {Muraveva}, {Musella},
  {Narbonne}, {Nelemans}, {Nicastro}, {Noval}, {Ord{\'e}novic},
  {Ordieres-Mer{\'e}}, {Osborne}, {Pagani}, {Pagano}, {Pailler}, {Palacin},
  {Palaversa}, {Parsons}, {Paulsen}, {Pecoraro}, {Pedrosa}, {Pentik{\"a}inen},
  {Pereira}, {Pichon}, {Piersimoni}, {Pineau}, {Plachy}, {Plum}, {Poujoulet},
  {Pr{\v{s}}a}, {Pulone}, {Ragaini}, {Rago}, {Rambaux}, {Ramos-Lerate},
  {Ranalli}, {Rauw}, {Read}, {Regibo}, {Renk}, {Reyl{\'e}}, {Ribeiro},
  {Rimoldini}, {Ripepi}, {Riva}, {Rixon}, {Roelens}, {Romero-G{\'o}mez},
  {Rowell}, {Royer}, {Rudolph}, {Ruiz-Dern}, {Sadowski}, {Sagrist{\`a}
  Sell{\'e}s}, {Sahlmann}, {Salgado}, {Salguero}, {Sarasso}, {Savietto},
  {Schnorhk}, {Schultheis}, {Sciacca}, {Segol}, {Segovia}, {Segransan},
  {Serpell}, {Shih}, {Smareglia}, {Smart}, {Smith}, {Solano}, {Solitro},
  {Sordo}, {Soria Nieto}, {Souchay}, {Spagna}, {Spoto}, {Stampa}, {Steele},
  {Steidelm{\"u}ller}, {Stephenson}, {Stoev}, {Suess}, {S{\"u}veges}, {Surdej},
  {Szabados}, {Szegedi-Elek}, {Tapiador}, {Taris}, {Tauran}, {Taylor},
  {Teixeira}, {Terrett}, {Tingley}, {Trager}, {Turon}, {Ulla}, {Utrilla},
  {Valentini}, {van Elteren}, {Van Hemelryck}, {van Leeuwen}, {Varadi},
  {Vecchiato}, {Veljanoski}, {Via}, {Vicente}, {Vogt}, {Voss}, {Votruba},
  {Voutsinas}, {Walmsley}, {Weiler}, {Weingrill}, {Werner}, {Wevers},
  {Whitehead}, {Wyrzykowski}, {Yoldas}, {{\v{Z}}erjal}, {Zucker}, {Zurbach},
  {Zwitter}, {Alecu}, {Allen}, {Allende Prieto}, {Amorim},
  {Anglada-Escud{\'e}}, {Arsenijevic}, {Azaz}, {Balm}, {Beck}, {Bernstein},
  {Bigot}, {Bijaoui}, {Blasco}, {Bonfigli}, {Bono}, {Boudreault}, {Bressan},
  {Brown}, {Brunet}, {Bunclark}, {Buonanno}, {Butkevich}, {Carret}, {Carrion},
  {Chemin}, {Ch{\'e}reau}, {Corcione}, {Darmigny}, {de Boer}, {de Teodoro}, {de
  Zeeuw}, {Delle Luche}, {Domingues}, {Dubath}, {Fodor}, {Fr{\'e}zouls},
  {Fries}, {Fustes}, {Fyfe}, {Gallardo}, {Gallegos}, {Gardiol}, {Gebran},
  {Gomboc}, {G{\'o}mez}, {Grux}, {Gueguen}, {Heyrovsky}, {Hoar}, {Iannicola},
  {Isasi Parache}, {Janotto}, {Joliet}, {Jonckheere}, {Keil}, {Kim},
  {Klagyivik}, {Klar}, {Knude}, {Kochukhov}, {Kolka}, {Kos}, {Kutka}, {Lainey},
  {LeBouquin}, {Liu}, {Loreggia}, {Makarov}, {Marseille}, {Martayan},
  {Martinez-Rubi}, {Massart}, {Meynadier}, {Mignot}, {Munari}, {Nguyen},
  {Nordlander}, {Ocvirk}, {O'Flaherty}, {Olias Sanz}, {Ortiz}, {Osorio},
  {Oszkiewicz}, {Ouzounis}, {Palmer}, {Park}, {Pasquato}, {Peltzer}, {Peralta},
  {P{\'e}turaud}, {Pieniluoma}, {Pigozzi}, {Poels}, {Prat}, {Prod'homme},
  {Raison}, {Rebordao}, {Risquez}, {Rocca-Volmerange}, {Rosen}, {Ruiz-Fuertes},
  {Russo}, {Sembay}, {Serraller Vizcaino}, {Short}, {Siebert}, {Silva},
  {Sinachopoulos}, {Slezak}, {Soffel}, {Sosnowska}, {Strai{\v{z}}ys}, {ter
  Linden}, {Terrell}, {Theil}, {Tiede}, {Troisi}, {Tsalmantza}, {Tur},
  {Vaccari}, {Vachier}, {Valles}, {Van Hamme}, {Veltz}, {Virtanen}, {Wallut},
  {Wichmann}, {Wilkinson}, {Ziaeepour}, \&
  {Zschocke}}]{gaia_collaboration_gaia_2016-1}
{Gaia Collaboration}, {Prusti}, T., {de Bruijne}, J.~H.~J., {et~al.}
  2016{\natexlab{b}}, \aap, 595, A1

\bibitem[{{Gaia Collaboration} {et~al.}(2023){Gaia Collaboration}, {Vallenari},
  {Brown}, {Prusti}, {de Bruijne}, {Arenou}, {Babusiaux}, {Biermann},
  {Creevey}, {Ducourant}, {Evans}, {Eyer}, {Guerra}, {Hutton}, {Jordi},
  {Klioner}, {Lammers}, {Lindegren}, {Luri}, {Mignard}, {Panem}, {Pourbaix},
  {Randich}, {Sartoretti}, {Soubiran}, {Tanga}, {Walton}, {Bailer-Jones},
  {Bastian}, {Drimmel}, {Jansen}, {Katz}, {Lattanzi}, {van Leeuwen}, {Bakker},
  {Cacciari}, {Casta{\~n}eda}, {De Angeli}, {Fabricius}, {Fouesneau},
  {Fr{\'e}mat}, {Galluccio}, {Guerrier}, {Heiter}, {Masana}, {Messineo},
  {Mowlavi}, {Nicolas}, {Nienartowicz}, {Pailler}, {Panuzzo}, {Riclet}, {Roux},
  {Seabroke}, {Sordo}, {Th{\'e}venin}, {Gracia-Abril}, {Portell}, {Teyssier},
  {Altmann}, {Andrae}, {Audard}, {Bellas-Velidis}, {Benson}, {Berthier},
  {Blomme}, {Burgess}, {Busonero}, {Busso}, {C{\'a}novas}, {Carry}, {Cellino},
  {Cheek}, {Clementini}, {Damerdji}, {Davidson}, {de Teodoro}, {Nu{\~n}ez
  Campos}, {Delchambre}, {Dell'Oro}, {Esquej}, {Fern{\'a}ndez-Hern{\'a}ndez},
  {Fraile}, {Garabato}, {Garc{\'\i}a-Lario}, {Gosset}, {Haigron}, {Halbwachs},
  {Hambly}, {Harrison}, {Hern{\'a}ndez}, {Hestroffer}, {Hodgkin}, {Holl},
  {Jan{\ss}en}, {Jevardat de Fombelle}, {Jordan}, {Krone-Martins}, {Lanzafame},
  {L{\"o}ffler}, {Marchal}, {Marrese}, {Moitinho}, {Muinonen}, {Osborne},
  {Pancino}, {Pauwels}, {Recio-Blanco}, {Reyl{\'e}}, {Riello}, {Rimoldini},
  {Roegiers}, {Rybizki}, {Sarro}, {Siopis}, {Smith}, {Sozzetti}, {Utrilla},
  {van Leeuwen}, {Abbas}, {{\'A}brah{\'a}m}, {Abreu Aramburu}, {Aerts},
  {Aguado}, {Ajaj}, {Aldea-Montero}, {Altavilla}, {{\'A}lvarez}, {Alves},
  {Anders}, {Anderson}, {Anglada Varela}, {Antoja}, {Baines}, {Baker},
  {Balaguer-N{\'u}{\~n}ez}, {Balbinot}, {Balog}, {Barache}, {Barbato},
  {Barros}, {Barstow}, {Bartolom{\'e}}, {Bassilana}, {Bauchet}, {Becciani},
  {Bellazzini}, {Berihuete}, {Bernet}, {Bertone}, {Bianchi}, {Binnenfeld},
  {Blanco-Cuaresma}, {Blazere}, {Boch}, {Bombrun}, {Bossini}, {Bouquillon},
  {Bragaglia}, {Bramante}, {Breedt}, {Bressan}, {Brouillet}, {Brugaletta},
  {Bucciarelli}, {Burlacu}, {Butkevich}, {Buzzi}, {Caffau}, {Cancelliere},
  {Cantat-Gaudin}, {Carballo}, {Carlucci}, {Carnerero}, {Carrasco},
  {Casamiquela}, {Castellani}, {Castro-Ginard}, {Chaoul}, {Charlot}, {Chemin},
  {Chiaramida}, {Chiavassa}, {Chornay}, {Comoretto}, {Contursi}, {Cooper},
  {Cornez}, {Cowell}, {Crifo}, {Cropper}, {Crosta}, {Crowley}, {Dafonte},
  {Dapergolas}, {David}, {David}, {de Laverny}, {De Luise}, {De March}, {De
  Ridder}, {de Souza}, {de Torres}, {del Peloso}, {del Pozo}, {Delbo},
  {Delgado}, {Delisle}, {Demouchy}, {Dharmawardena}, {Di Matteo}, {Diakite},
  {Diener}, {Distefano}, {Dolding}, {Edvardsson}, {Enke}, {Fabre}, {Fabrizio},
  {Faigler}, {Fedorets}, {Fernique}, {Fienga}, {Figueras}, {Fournier},
  {Fouron}, {Fragkoudi}, {Gai}, {Garcia-Gutierrez}, {Garcia-Reinaldos},
  {Garc{\'\i}a-Torres}, {Garofalo}, {Gavel}, {Gavras}, {Gerlach}, {Geyer},
  {Giacobbe}, {Gilmore}, {Girona}, {Giuffrida}, {Gomel}, {Gomez},
  {Gonz{\'a}lez-N{\'u}{\~n}ez}, {Gonz{\'a}lez-Santamar{\'\i}a},
  {Gonz{\'a}lez-Vidal}, {Granvik}, {Guillout}, {Guiraud},
  {Guti{\'e}rrez-S{\'a}nchez}, {Guy}, {Hatzidimitriou}, {Hauser}, {Haywood},
  {Helmer}, {Helmi}, {Sarmiento}, {Hidalgo}, {Hilger}, {H{\l}adczuk}, {Hobbs},
  {Holland}, {Huckle}, {Jardine}, {Jasniewicz}, {Jean-Antoine Piccolo},
  {Jim{\'e}nez-Arranz}, {Jorissen}, {Juaristi Campillo}, {Julbe}, {Karbevska},
  {Kervella}, {Khanna}, {Kontizas}, {Kordopatis}, {Korn}, {K{\'o}sp{\'a}l},
  {Kostrzewa-Rutkowska}, {Kruszy{\'n}ska}, {Kun}, {Laizeau}, {Lambert},
  {Lanza}, {Lasne}, {Le Campion}, {Lebreton}, {Lebzelter}, {Leccia}, {Leclerc},
  {Lecoeur-Taibi}, {Liao}, {Licata}, {Lindstr{\o}m}, {Lister}, {Livanou},
  {Lobel}, {Lorca}, {Loup}, {Madrero Pardo}, {Magdaleno Romeo}, {Managau},
  {Mann}, {Manteiga}, {Marchant}, {Marconi}, {Marcos}, {Marcos Santos},
  {Mar{\'\i}n Pina}, {Marinoni}, {Marocco}, {Marshall}, {Martin Polo},
  {Mart{\'\i}n-Fleitas}, {Marton}, {Mary}, {Masip}, {Massari},
  {Mastrobuono-Battisti}, {Mazeh}, {McMillan}, {Messina}, {Michalik}, {Millar},
  {Mints}, {Molina}, {Molinaro}, {Moln{\'a}r}, {Monari}, {Mongui{\'o}},
  {Montegriffo}, {Montero}, {Mor}, {Mora}, {Morbidelli}, {Morel}, {Morris},
  {Muraveva}, {Murphy}, {Musella}, {Nagy}, {Noval}, {Oca{\~n}a}, {Ogden},
  {Ordenovic}, {Osinde}, {Pagani}, {Pagano}, {Palaversa}, {Palicio},
  {Pallas-Quintela}, {Panahi}, {Payne-Wardenaar}, {Pe{\~n}alosa Esteller},
  {Penttil{\"a}}, {Pichon}, {Piersimoni}, {Pineau}, {Plachy}, {Plum}, {Poggio},
  {Pr{\v{s}}a}, {Pulone}, {Racero}, {Ragaini}, {Rainer}, {Raiteri}, {Rambaux},
  {Ramos}, {Ramos-Lerate}, {Re Fiorentin}, {Regibo}, {Richards}, {Rios Diaz},
  {Ripepi}, {Riva}, {Rix}, {Rixon}, {Robichon}, {Robin}, {Robin}, {Roelens},
  {Rogues}, {Rohrbasser}, {Romero-G{\'o}mez}, {Rowell}, {Royer}, {Ruz Mieres},
  {Rybicki}, {Sadowski}, {S{\'a}ez N{\'u}{\~n}ez}, {Sagrist{\`a} Sell{\'e}s},
  {Sahlmann}, {Salguero}, {Samaras}, {Sanchez Gimenez}, {Sanna},
  {Santove{\~n}a}, {Sarasso}, {Schultheis}, {Sciacca}, {Segol}, {Segovia},
  {S{\'e}gransan}, {Semeux}, {Shahaf}, {Siddiqui}, {Siebert}, {Siltala},
  {Silvelo}, {Slezak}, {Slezak}, {Smart}, {Snaith}, {Solano}, {Solitro},
  {Souami}, {Souchay}, {Spagna}, {Spina}, {Spoto}, {Steele},
  {Steidelm{\"u}ller}, {Stephenson}, {S{\"u}veges}, {Surdej}, {Szabados},
  {Szegedi-Elek}, {Taris}, {Taylor}, {Teixeira}, {Tolomei}, {Tonello}, {Torra},
  {Torra}, {Torralba Elipe}, {Trabucchi}, {Tsounis}, {Turon}, {Ulla}, {Unger},
  {Vaillant}, {van Dillen}, {van Reeven}, {Vanel}, {Vecchiato}, {Viala},
  {Vicente}, {Voutsinas}, {Weiler}, {Wevers}, {Wyrzykowski}, {Yoldas}, {Yvard},
  {Zhao}, {Zorec}, {Zucker}, \& {Zwitter}}]{gaia_collaboration_gaia_2023}
{Gaia Collaboration}, {Vallenari}, A., {Brown}, A.~G.~A., {et~al.} 2023, \aap,
  674, A1

\bibitem[{{Grand} {et~al.}(2023){Grand}, {Pakmor}, {Fragkoudi}, {G{\'o}mez},
  {Trick}, {Simpson}, {van de Voort}, \& {Bieri}}]{grand_ever-present_2023}
{Grand}, R. J.~J., {Pakmor}, R., {Fragkoudi}, F., {et~al.} 2023, \mnras, 524,
  801

\bibitem[{{GRAVITY Collaboration} {et~al.}(2018){GRAVITY Collaboration},
  {Abuter}, {Amorim}, {Anugu}, {Baub{\"o}ck}, {Benisty}, {Berger}, {Blind},
  {Bonnet}, {Brandner}, {Buron}, {Collin}, {Chapron}, {Cl{\'e}net}, {Coud{\'e}
  Du Foresto}, {de Zeeuw}, {Deen}, {Delplancke-Str{\"o}bele}, {Dembet},
  {Dexter}, {Duvert}, {Eckart}, {Eisenhauer}, {Finger}, {F{\"o}rster
  Schreiber}, {F{\'e}dou}, {Garcia}, {Garcia Lopez}, {Gao}, {Gendron},
  {Genzel}, {Gillessen}, {Gordo}, {Habibi}, {Haubois}, {Haug}, {Hau{\ss}mann},
  {Henning}, {Hippler}, {Horrobin}, {Hubert}, {Hubin}, {Jimenez Rosales},
  {Jochum}, {Jocou}, {Kaufer}, {Kellner}, {Kendrew}, {Kervella}, {Kok},
  {Kulas}, {Lacour}, {Lapeyr{\`e}re}, {Lazareff}, {Le Bouquin}, {L{\'e}na},
  {Lippa}, {Lenzen}, {M{\'e}rand}, {M{\"u}ler}, {Neumann}, {Ott}, {Palanca},
  {Paumard}, {Pasquini}, {Perraut}, {Perrin}, {Pfuhl}, {Plewa}, {Rabien},
  {Ram{\'\i}rez}, {Ramos}, {Rau}, {Rodr{\'\i}guez-Coira}, {Rohloff}, {Rousset},
  {Sanchez-Bermudez}, {Scheithauer}, {Sch{\"o}ller}, {Schuler}, {Spyromilio},
  {Straub}, {Straubmeier}, {Sturm}, {Tacconi}, {Tristram}, {Vincent}, {von
  Fellenberg}, {Wank}, {Waisberg}, {Widmann}, {Wieprecht}, {Wiest},
  {Wiezorrek}, {Woillez}, {Yazici}, {Ziegler}, \&
  {Zins}}]{gravity_collaboration_detection_2018}
{GRAVITY Collaboration}, {Abuter}, R., {Amorim}, A., {et~al.} 2018, \aap, 615,
  L15

\bibitem[{{Guo} {et~al.}(2024){Guo}, {Li}, {Shen}, {Mao}, \&
  {Liu}}]{guo_measuring_2024}
{Guo}, R., {Li}, Z.-Y., {Shen}, J., {Mao}, S., \& {Liu}, C. 2024, \apj, 960,
  133

\bibitem[{{Harris} {et~al.}(2020){Harris}, Millman, van~der Walt, Gommers,
  Virtanen, Cournapeau, Wieser, Taylor, Berg, Smith, Kern, Picus, Hoyer, van
  Kerkwijk, Brett, Haldane, del Río, Wiebe, Peterson, Gérard-Marchant,
  Sheppard, Reddy, Weckesser, Abbasi, Gohlke, \& Oliphant}]{harris_array_2020}
{Harris}, C.~R., Millman, K.~J., van~der Walt, S.~J., {et~al.} 2020, Nature,
  585, 357

\bibitem[{{Hunt} {et~al.}(2022){Hunt}, {Price-Whelan}, {Johnston}, \&
  {Darragh-Ford}}]{hunt_multiple_2022}
{Hunt}, J. A.~S., {Price-Whelan}, A.~M., {Johnston}, K.~V., \& {Darragh-Ford},
  E. 2022, \mnras, 516, L7

\bibitem[{{Hunt} {et~al.}(2024){Hunt}, {Price-Whelan}, {Johnston}, {McClure},
  {Filion}, {Cassese}, \& {Horta}}]{hunt_radial_2024}
{Hunt}, J. A.~S., {Price-Whelan}, A.~M., {Johnston}, K.~V., {et~al.} 2024,
  \mnras, 527, 11393

\bibitem[{{Hunt} {et~al.}(2021){Hunt}, {Stelea}, {Johnston}, {Gandhi},
  {Laporte}, \& {B{\'e}dorf}}]{hunt_resolving_2021}
{Hunt}, J. A.~S., {Stelea}, I.~A., {Johnston}, K.~V., {et~al.} 2021, \mnras,
  508, 1459

\bibitem[{Hunter(2007)}]{hunter_matplotlib_2007}
Hunter, J.~D. 2007, Computing in Science \& Engineering, 9, 90

\bibitem[{{Kerr}(1957)}]{kerr_magellanic_1957}
{Kerr}, F.~J. 1957, \aj, 62, 93

\bibitem[{{Laporte} {et~al.}(2019){Laporte}, {Minchev}, {Johnston}, \&
  {G{\'o}mez}}]{laporte_footprints_2019}
{Laporte}, C. F.~P., {Minchev}, I., {Johnston}, K.~V., \& {G{\'o}mez}, F.~A.
  2019, \mnras, 485, 3134

\bibitem[{{Li} {et~al.}(2023){Li}, {Siebert}, {Monari}, {Famaey}, \&
  {Rozier}}]{li_gaia_2023}
{Li}, C., {Siebert}, A., {Monari}, G., {Famaey}, B., \& {Rozier}, S. 2023,
  \mnras, 524, 6331

\bibitem[{{Li}(2021)}]{li_vertical_2021}
{Li}, Z.-Y. 2021, \apj, 911, 107

\bibitem[{{Lindblad}(1941)}]{lindblad_development_1941}
{Lindblad}, B. 1941, Stockholms Observatoriums Annaler, 13, 10.1

\bibitem[{{McMillan}(2017)}]{mcmillan_mass_2017}
{McMillan}, P.~J. 2017, \mnras, 465, 76

\bibitem[{{Poggio} {et~al.}(2018){Poggio}, {Drimmel}, {Lattanzi}, {Smart},
  {Spagna}, {Andrae}, {Bailer-Jones}, {Fouesneau}, {Antoja}, {Babusiaux},
  {Evans}, {Figueras}, {Katz}, {Reyl{\'e}}, {Robin}, {Romero-G{\'o}mez}, \&
  {Seabroke}}]{poggio_galactic_2018}
{Poggio}, E., {Drimmel}, R., {Lattanzi}, M.~G., {et~al.} 2018, \mnras, 481, L21

\bibitem[{{Ramos} {et~al.}(2018){Ramos}, {Antoja}, \&
  {Figueras}}]{ramos_riding_2018}
{Ramos}, P., {Antoja}, T., \& {Figueras}, F. 2018, \aap, 619, A72

\bibitem[{{Recio-Blanco} {et~al.}(2023){Recio-Blanco}, {de Laverny}, {Palicio},
  {Kordopatis}, {{\'A}lvarez}, {Schultheis}, {Contursi}, {Zhao}, {Torralba
  Elipe}, {Ordenovic}, {Manteiga}, {Dafonte}, {Oreshina-Slezak}, {Bijaoui},
  {Fr{\'e}mat}, {Seabroke}, {Pailler}, {Spitoni}, {Poggio}, {Creevey}, {Abreu
  Aramburu}, {Accart}, {Andrae}, {Bailer-Jones}, {Bellas-Velidis}, {Brouillet},
  {Brugaletta}, {Burlacu}, {Carballo}, {Casamiquela}, {Chiavassa}, {Cooper},
  {Dapergolas}, {Delchambre}, {Dharmawardena}, {Drimmel}, {Edvardsson},
  {Fouesneau}, {Garabato}, {Garc{\'\i}a-Lario}, {Garc{\'\i}a-Torres}, {Gavel},
  {Gomez}, {Gonz{\'a}lez-Santamar{\'\i}a}, {Hatzidimitriou}, {Heiter},
  {Jean-Antoine Piccolo}, {Kontizas}, {Korn}, {Lanzafame}, {Lebreton}, {Le
  Fustec}, {Licata}, {Lindstr{\o}m}, {Livanou}, {Lobel}, {Lorca}, {Magdaleno
  Romeo}, {Marocco}, {Marshall}, {Mary}, {Nicolas}, {Pallas-Quintela}, {Panem},
  {Pichon}, {Riclet}, {Robin}, {Rybizki}, {Santove{\~n}a}, {Silvelo}, {Smart},
  {Sarro}, {Sordo}, {Soubiran}, {S{\"u}veges}, {Ulla}, {Vallenari}, {Zorec},
  {Utrilla}, \& {Bakker}}]{recio-blanco_gaia_2023}
{Recio-Blanco}, A., {de Laverny}, P., {Palicio}, P.~A., {et~al.} 2023, \aap,
  674, A29

\bibitem[{{Reid} \& {Brunthaler}(2004)}]{reid_proper_2004}
{Reid}, M.~J. \& {Brunthaler}, A. 2004, \apj, 616, 872

\bibitem[{{Sellwood} \& {Binney}(2002)}]{sellwood_radial_2002}
{Sellwood}, J.~A. \& {Binney}, J.~J. 2002, \mnras, 336, 785

\bibitem[{{Tremaine} {et~al.}(2023){Tremaine}, {Frankel}, \&
  {Bovy}}]{tremaine_origin_2023}
{Tremaine}, S., {Frankel}, N., \& {Bovy}, J. 2023, \mnras, 521, 114

\bibitem[{{Vasiliev}(2019)}]{vasiliev_agama_2019}
{Vasiliev}, E. 2019, \mnras, 482, 1525

\bibitem[{{Virtanen} {et~al.}(2020){Virtanen}, {Gommers}, {Oliphant},
  {Haberland}, {Reddy}, {Cournapeau}, {Burovski}, {Peterson}, {Weckesser},
  {Bright}, {van der Walt}, {Brett}, {Wilson}, {Millman}, {Mayorov}, {Nelson},
  {Jones}, {Kern}, {Larson}, {Carey}, {Polat}, {Feng}, {Moore}, {VanderPlas},
  {Laxalde}, {Perktold}, {Cimrman}, {Henriksen}, {Quintero}, {Harris},
  {Archibald}, {Ribeiro}, {Pedregosa}, {van Mulbregt}, \& {SciPy 1. 0
  Contributors}}]{virtanen_scipy_2020}
{Virtanen}, P., {Gommers}, R., {Oliphant}, T.~E., {et~al.} 2020, Nature
  Methods, 17, 261

\bibitem[{{White} \& {Rees}(1978)}]{white_core_1978}
{White}, S.~D.~M. \& {Rees}, M.~J. 1978, \mnras, 183, 341

\bibitem[{{Widmark} {et~al.}(2021){Widmark}, {Laporte}, \& {de
  Salas}}]{widmark_weighing_2021-1}
{Widmark}, A., {Laporte}, C., \& {de Salas}, P.~F. 2021, \aap, 650, A124

\bibitem[{{Widmark} {et~al.}(2022){Widmark}, {Laporte}, \&
  {Monari}}]{widmark_weighing_2022-1}
{Widmark}, A., {Laporte}, C.~F.~P., \& {Monari}, G. 2022, \aap, 663, A15

\bibitem[{{Widrow}(2023)}]{widrow_swing_2023}
{Widrow}, L.~M. 2023, \mnras, 522, 477

\end{thebibliography}

\end{document}